# Preserve or retreat? Willingness-to-pay for Coastline Protection in New South Wales

**Ardeshiri A.[1,2,*], Swait J.[2], Heagney E.C.[3,4], Kovac M.[3]**

1- Research Centre for Integrated Transport Innovation, School of Civil and Environmental Engineering, UNSW Sydney, NSW 2052, Australia

2- Institute for Choice, University of South Australia, 140 Arthur St, North Sydney, NSW 2060

3- NSW Office of Environment and Heritage, PO Box A290, Sydney South, NSW, 1232

4- School of Environment, Science and Engineering, Southern Cross University, Military Rd, Lismore, NSW 2480.

* Corresponding author – email: A.Ardeshiri@unsw.edu.au

Type of paper: *Research paper*

Declarations of interest: *none*

## ABSTRACT

Coastal erosion is a global and pervasive phenomenon that predicates a need for a strategic approach to the future management of coastal values and assets (both built and natural): should we invest in protective structures like seawalls that aim to preserve specific coastal features, or allow natural coastline retreat to preserve sandy beaches and other coastal ecosystems? Determining the most suitable management approach in a specific context requires a better understanding of the full suite of economic values the populations holds for coastal assets, including non-market values. In this study, we characterise New South Wales residents' willingness to pay to maintain sandy beaches (width and length) in the face of coastal erosion along the Australian state's nearly 2200 km coastline. The measurement instrument is a stated preference referendum task administered state-wide to a sample of 2014 respondents, with the payment mechanism defined as a purpose-specific incremental levy of a fixed amount over a set period of years. We use an innovative application of a Latent Class Binary Logit model to deal with "Yea-sayers" and "Nay-sayers", as well as revealing the latent heterogeneity among sample members. We find that 65% of the population would be willing to pay some amount of levy, dependent on the policy setting. In most cases, there is no effect of degree of beach deterioration – characterised as loss of width and/or length of sandy beaches of between 5% and 100% - on respondents' willingness to pay for a management levy. This suggests that respondents who agreed to pay a management levy were motivated to preserve sandy beaches in their current state irrespective of the severity of sand loss likely to occur as a result of coastal erosion. Willingness to pay also varies according to beach type (amongst Iconic, Main, Bay and Surf beaches) – a finding that can assist with spatial prioritisation of coastal management. Not recognizing the presence of nay-sayers in the data or recognizing them but eliminating them from the estimation will result in biased WTP results and, consequently, biased policy propositions by coastal managers.

**KEYWORDS:** Referendum, Coastal Management, Choice Modelling, Beach Erosion, Levy, Protest voting

# 1. INTRODUCTION

Beach environments provide various services such as aesthetic beauty, habitat for marine and terrestrial plants and animals, transportation, protection from coastal hazards, opportunities for recreation and income generation (Brenner, Jiménez, Sardá, & Garola, 2010; Camacho-Valdez, Ruiz-Luna, Ghermandi, & Nunes, 2013; Windle & Rolfe, 2013). This array of services has accelerated industrialization and urbanization processes along the coast and given rise to coastal population centres and coast-dependent economies (Falco, 2017). Changes in climate have also contributed to changes in coastal communities over decadal and millennial timeframes (Short & Woodroffe, 2009; Vörösmarty, Green, Salisbury, & Lammers, 2000). Natural and man-made processes continue to shape coastal environments, including through their contribution to coastal erosion. The literature identifies coastal erosion as a global process and a significant challenge for effective management of coastal values into the future (Norman 2009, Caton & Harvey, 2010, Windle & Rolfe, 2013).

Coastal erosion demands an ongoing need for management of both natural coastal assets and the built assets that have accumulated in the coastal zone. Coastal erosion can introduce or increase competition amongst land-uses or asset classes (Phillips & Jones 2006, Titus et al. 2009). As coastlines erode, land may be assigned to build assets like roads, houses or infrastructure, or to natural assets like beaches. Each of these assets types will provide a different level of economic value to the surrounding community, so the economic implications of asset trade-offs need to be considered in management decisions.

Primary management options include a) preserving specific coastal assets (usually built assets) by constructing natural or engineering buffers that limit erosion of specific sites deemed to be of high value (Abel et al. 2011), or b) undertaking "planned retreat", whereby any built assets deemed to be at risk of erosion are systematically and sequentially removed from the erosion zone. The planned retreat strategy focusses on maintaining natural assets, including sandy beaches, by removing any hard infrastructure surfaces that might otherwise act as barriers to their natural landward migration; these natural assets subsequently act as natural barriers, protecting remaining built infrastructure from further damage (Doody 2004, McGranahan et al. 2007). These alternate management strategies highlight the difficult trade-offs between natural and built assets that are inherent in managing coastal erosion. In reality, management options can be less clear-cut. For example, seawalls or other engineering structures may be built to protect natural assets (like specific surf breaks that are deemed to be of special significance or which provide a large economic return from tourism); engineering works may be augmented with 'beach nourishment' to try to maintain both built and natural values at a given location.

Coastal protection in Australia has traditionally focussed on the protection of built assets through the use of sea-walls or other engineering structures without considering the economic implications of alternative land allocation strategies (Gurran et al. 2007). This has been achieved through the allocation of public funds by local and state governments alone or in partnership with the federal government (Gurran et al. 2007, McFadden 2010). In some cases, illegal coastal protection works have been carried out by individuals to protect private assets (usually homes). Whilst these strategies have been successful in protecting built assets at high-risk coastal locations, they can also have adverse implications for other coastal values. Coastal protection works can also come as a more direct (in-place) trade-off with sandy beaches or other coastal ecosystems like mangroves and saltmarsh because the preservation of hard infrastructure makes the landward migration of these ecosystems impossible (Nordstrom 2003, Feagin et al. 2005). Sandy beach loss or narrowing resulting from seawalls or other protective structures has been reported in Australia (Abel et al. 2011) and at a large number of locations around the globe (Fletcher et al. 1997, Phillips & Jones 2006). The value of sandy beach (and other coastal ecosystems) losses are rarely accounted for when the costs and benefits of coastal protection are being assessed (Phillips & Jones 2006, Abel et al. 2011). Moreover, protective strategies tend to occur in an ad-hoc or reactive manner and focus on mitigating losses for stakeholders who are directly affected by erosion (e.g. directly impacted homeowners) without necessarily accounting for the broader suite of values, like recreation or non-use values, that the population might hold for other assets (Abel et al. 2011).

Different coastal erosion management strategies are likely to be relevant at different coastal locations dependent on community preferences for the configuration of the coastal zone, including the appropriate mix of built versus natural assets, into the future. Irrespective of the management option that is ultimately selected in a given setting, securing coastal assets for future use will require a significant investment in long-term planning and management. To ensure effective and efficient future management it will be necessary to a) enhance their financial sustainability to ensure the level of funding allocated to coastal erosion management is sufficient for management costs, b) ensure that management is in line with community values and preferences for future coastal configurations and c) move away from ad-hoc protection and repair works towards a more strategic management approach that prioritises the maintenance and protection of natural and/or built assets at key priority locations. Multiple questions then arise … Are citizens willing to invest in the maintenance of their coast? How does management account for affected parties (like homeowners) as well as other stakeholders, like those who use coastal sites for recreation, or those who place a high value on the protection of coastal ecosystems? Which beach(es) should receive higher funding priority?

Addressing these questions requires a better understanding of the full suite of economic values the population holds for coastal assets. In this context, it is information relating to the non-use values of coastal areas that are currently most lacking. These non-use values encompass existence value –the value associated with knowing that biodiversity and other environmental values continue to exist (Perace & Moran 2013), bequest value – "a willingness to pay to preserve the environment for the benefit of one's descendants" (Turner et al. 1994) or for future generations (Pearce & Moran 2013), and option value – a value people place on potential future use of an environmental site or resource (Stevens et al. 1991). These are best quantified using non-market stated preference techniques like Contingent Valuation and Choice Modelling (see Methods).

This study seeks to quantify non-use values for a specific coastal asset type – sandy beaches – held by households along the coast of New South Wales in south-eastern Australia. We have designed this study so that it will address a number of the management challenges identified above. First, the payment mechanism employed is a targeted incremental annual levy. This is a common payment mechanism in Australia and provides managers with an estimate of residents' willingness to pay for management that prevents beach loss, as well as a realistic vehicle through which sustainable financing of coastal management might be achieved. Second, we employ a repeated, hypothetical referendum task to compare willingness to pay (WTP) among four beach categories (Surf beaches, Bay beaches, Main beaches and Iconic beaches) and in response to the travel distance between a specific beach location and the respondent's home. These elements of our survey design go to addressing the question of how investment should be prioritised amongst a range of beaches along a given coastline - in this case, amongst the 755 open coast beaches that are exposed to erosion processes along the NSW coast.

From an applied point of view, this study contributes to the current literature by providing significant empirical findings that coastal managers can benefit from in their decision making about how to sustainably and efficiently finance coastal management into the future. From a methodological perspective, this study uses an innovative latent class model to infer both preference heterogeneity and to identify and deal with strategic (or protest) voting in the form of "nay-" as well as "yea-saying" during estimation. We present an alternative approach for dealing with nay- and yea-saying; many studies exclude these groups from overall estimates of WTP, which we argue can lead to a serious under- or over-estimation of value.

The remainder of the article proceeds as follows. First, we review some of the relevant literature. We follow with a description of the method and data used for the study. In the penultimate section, we report results of residents' willingness to pay to maintain sandy beaches in the face of coastal erosion pressures. We conclude with a discussion of the policy and research implications of our findings.

# 2. BACKGROUND AND CASE STUDY CONTEXT

The study was conducted in the state of New South Wales (NSW) in Australia. NSW has a total coastline length of 2194 km, equivalent to 3.6% of Australia's total coastline (Australia, 2004). The NSW coast is a dynamic place and since Australia's initial human occupation over 50,000 years ago, people have witnessed major changes in sea level, habitats and shape of the shoreline from great storm events. Over the geological past this dynamism has been even more pronounced, with sea levels up to 4–6 metres higher than today and the shoreline in some places more than 500 kilometres inland (Australian Government - Department of Climate Change, 2009). Research, presented at the Copenhagen Climate Congress in March 2009 projected sea-level rise from 75 centimetres to 190 centimetres relative to 1990, with 110–120 centimetres the mid-range of the projection. Based on this projection, in 2011 the Australian Government selected 1.1 metres as a plausible value for sea-level rise by high-end scenario for 2100 (Australian Government - Department of Climate Change, 2011). It is expected that up to $63 billion of existing residential buildings are potentially at risk of inundation from a 1.1 metre sea-level rise, with a lower and upper estimate of risk identified for between 157,000 and 247,600 individual buildings (refer to Figure 1 for estimated number of existing residential buildings at risk for the different Australian States).

**Figure 1:** Estimated number of existing residential buildings at risk of inundation from 1.1 meter sea-level rise (incl. 1-in- 100 storm tide for NSW, VIC and TAS and high tide event for others).

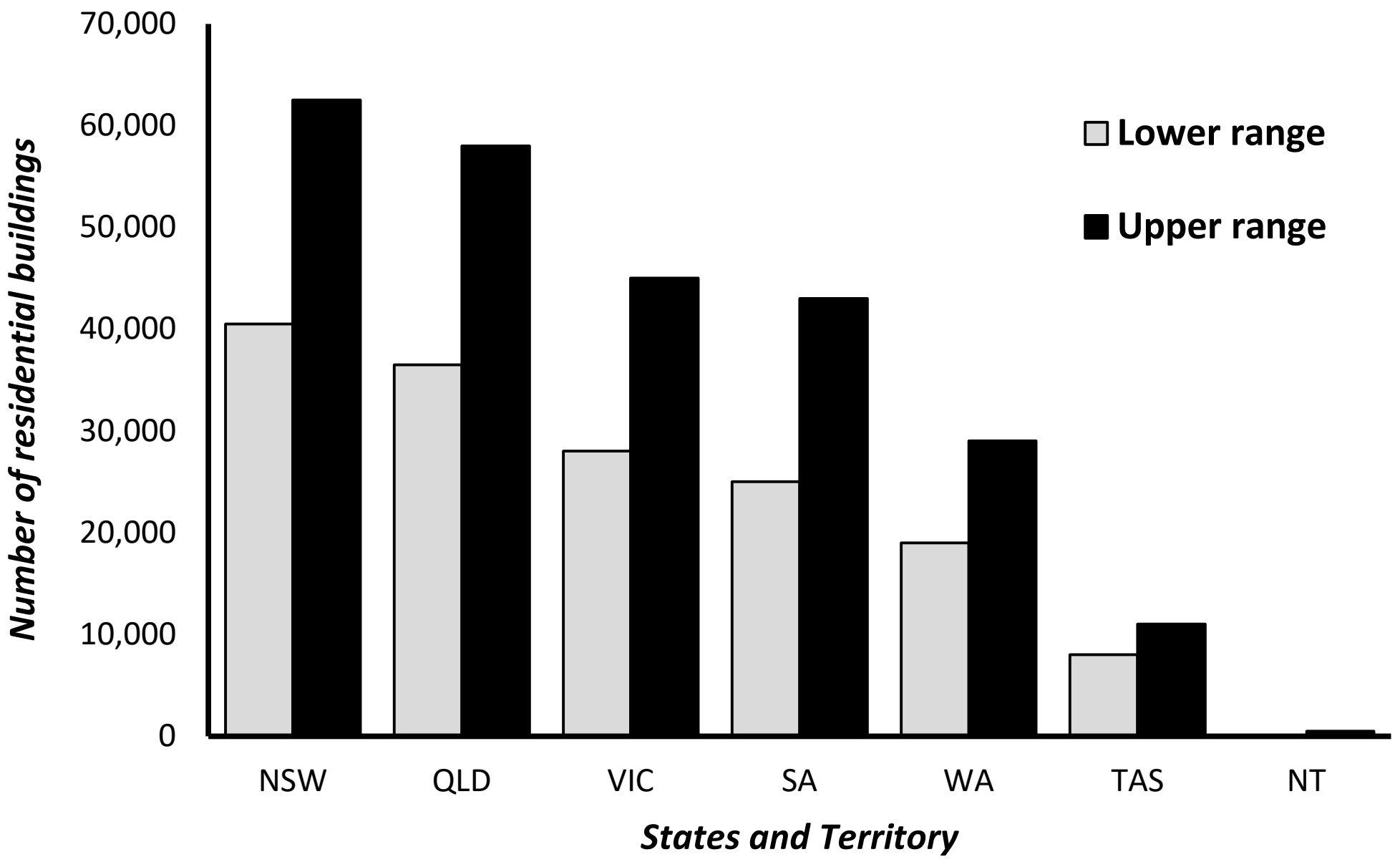

**Source:** https://www.environment.gov.au/system/files/resources/fa553e97-2ead-47bb-ac80-c12adffea944/files/cc-risks-full-report.pdf p.7

Severe episodic coastal erosion, particularly associated with a sequence of East Coast lows, has also occurred at several points along the New South Wales coast. The 1974 storms, estimated as a 1 in 200 years event, destroyed the pier at Manly and resulted in loss and damage to property, and roads being cut at several points in Sydney and at Moruya, Bermagui and Tathra. Elsewhere, this was the first stage of erosion, with subsequent storms actually undermining property, as with the three houses that were destroyed during a storm in 1978 at Wamberal. Short & Woodroffe (2009) describe Sydney's Collaroy beach as a classic example of inappropriate planning and shoreline subdivision that took place more than 100 years ago. The original property boundaries extended, and still do, down across the dune and onto the beach, with most of the houses and now some high-rise dwellings built on the beach-dune area. Long-term monitoring suggests that the beach has not receded over time (Harley et al. 2011), but infrastructure has incurred significant damage in major erosion events in 1920, 1944-45, 1967, 1974, and most recently in 2016. Figure 2, presents images of some major NSW erosion events. Similar coastal erosion events have happened in other States (i.e. Gold Coast 1967, and Victoria 2007).

**Figure 2 :** Examples of Erosion Events in NSW

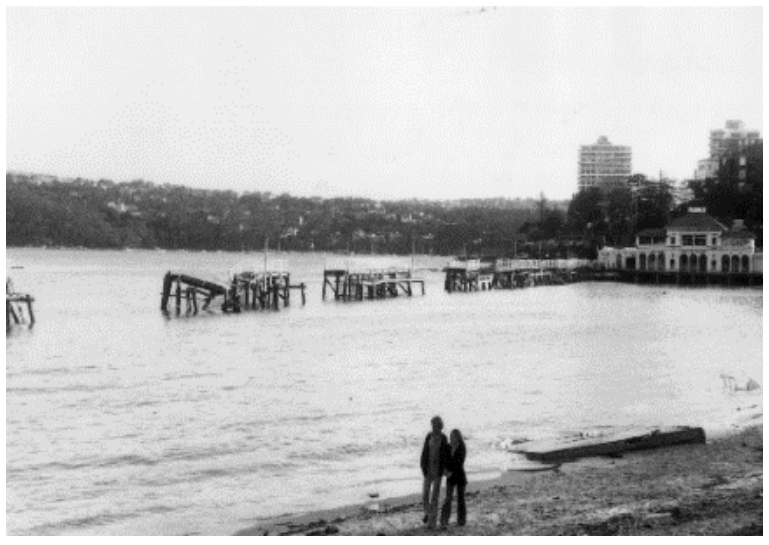

1974 - Damaged boardwalk, Manly Cove[1]

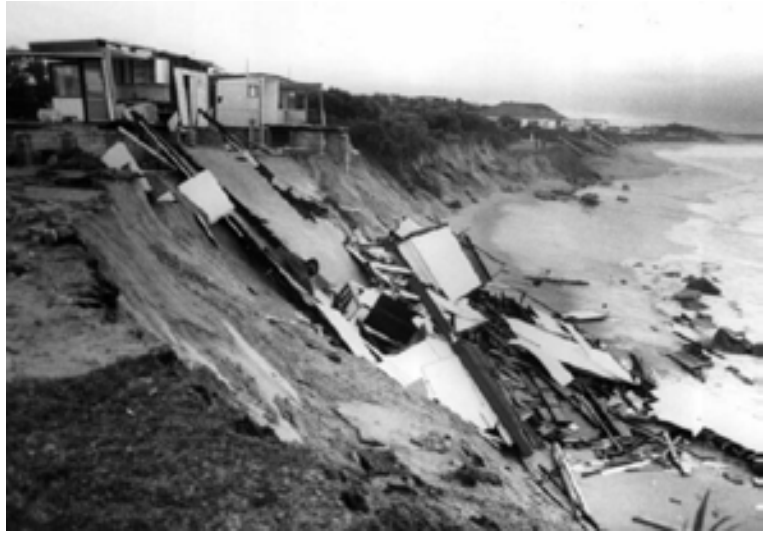

*1978 – Houses at Wamberal beach*[2]

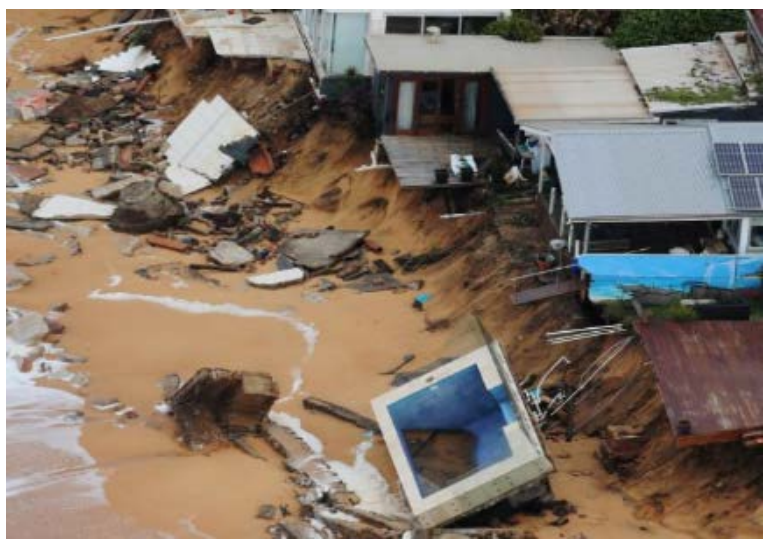

*2016 – Houses at Collaroy beach*[3]

In Australia, the responsibility for management of coastal areas is shared between three levels of government: Commonwealth (Federal), State and Local. The Commonwealth government lacks direct constitutional power in coastal management, while State coastal strategies generally outline the principles for coastal planning and management (Cooke, Jones, Goodwin, & Bishop, 2012; James, 2000). Local Government Authorities (LGAs) are responsible for implementing the majority of coastal management, translating State planning and management policies and legislation into local actions as well as providing infrastructure, foreshore maintenance and ensuring public safety (Cooke, et al., 2012; James, 2000). There are 755 open coast beaches in NSW which are exposed to a persistent, moderate southeast swell and tides of less than 2 meters. These beaches have an average beach length of 1.35 Kilometres, with the longest ("Ten Mile") 26 kilometres long (Short & Woodroffe, 2009).

For this study beaches where categorised into four classifications to allow generalization of results to any of the 755 beaches. These classifications are as follow[4]:

1. **Iconic beaches:** An Iconic beach is a well-known or famous beach that has a high visitation by users coming from outside the local council area.
2. **Surf beaches:** A Surf beach is an open coast beach with a surf break i.e.- a permanent obstruction such as a coral reef, rock, shoal, or headland that causes waves to break.

[1] Source: Water Research Laboratory, School of Civil and Environmental Engineering, UNSW Sydney

[2] Photo by Andr*ew Shot*: http://www.dailytelegraph.com.au/newslocal/central-coast/major-erosion-threatens-400m-worth-of-wamberal-homes-infrastructure/news-story/9897149515c200b5f8d2fc1e4a7f38f1

[3] Sources: http://www.news.com.au/technology/environment/collaroy-mansions-survive-overnight-king-tide-but-clean-up-causes-conflict/news-story/b60aaf9aa4bd7309a1b905559ff06d26

[4] The categorisation of the beaches was developed in consultation with the NSW Office of Environment and Heritage.

3. **Sheltered/Bay beaches:** A Sheltered/Bay beach is a beach that is protected from intense wave action because of a reef or rock formation that breaks the surf before it enters the Bay.
4. **Main beaches:** This refers to the beach located near a town that most people may visit as opposed to other beaches in the Local Council area that are less visited.

# 3. MATERIALS AND METHODS

## 3.1 THE CHOICE EXPERIMENT

Over two decades ago, the National Oceanic and Atmospheric Administration (NOAA) Blue Ribbon Panel report on Contingent Valuation (CV) stimulated a research agenda that fundamentally influenced the design and conduct of Stated Preference (SP) studies, particularly within the context of environmental valuation (Arrow, et al., 1993). The Panel focused on the use of CV to estimate non-use values for litigation in the United States and proposed what they referred to as "a fairly complete set of guidelines compliance with which would define an ideal CV survey." (Arrow et al., 1993; pp. 29) These guidelines spurred research to advance the validity and reliability of CV methods and were an indirect impetus for the expanding use of Choice Experiments (CEs).

SP has advanced considerably since the NOAA panel and this evolving research affects the applicability of the NOAA panel's guidelines. Referendum Choice Experiments (RCE) have become widely used as a technique for eliciting the value of public goods or non-market resources (Cameron, 1988; Green, Jacowitz, Kahneman, & McFadden, 1998; Johnston, et al., 2017). Subjects are presented with a hypothetical referendum that specifies a good to be supplied and a payment and asked to vote on this referendum. The payment, or *bid*, is varied experimentally to provide a profile of the cumulative distribution function of WTP at the experimental design points. This protocol has gained widespread use in the valuation of natural resources and has largely displaced older protocols in which subjects were asked to state an open-ended WTP for a good or to reveal a WTP range by responses to a sequence of bids or choices from a set of alternatives. However, practitioners have found that responses are influenced by the payment vehicle. This may arise from incentive effects of the 'free-rider' variety, or from the concerns of subjects about distributional implications and 'fairness' (Green, Jacowitz, Kahneman, & McFadden, 1998).

The concepts in economic theory underlying referendum surveys are preferences characterized in monetary units (*consumer surplus, compensating variation, willingness to pay*), the *Kaldor–Hicks compensation principle* as a criterion for aggregating individual preferences into a social choice rule, and Samuelson's theory of optimal supply of public goods, developed in a stream of literature that has emphasized *incentive-compatible mechanisms* that blunt the 'free-rider' problem (Green, et al., 1998). To be incentive compatible, a referendum on a pure public good needs to be a "*take-it*" or "*leave-it*" offer, where the vote doesn't influence any other offers that may be made to agents and where the payment mechanism is coercive in the sense that each agent can be required to pay independently of how the individual agent voted (Carson & Czajkowski, 2014). Many economists believe that if subjects are adequately economically motivated, the cognitive paradoxes sometimes observed in psychological experiments disappear (Green et al., 1998). Thus, a decision rule should be selected that is realistic and binding on respondents. In Australian (and NSW) political settings direct democracy is practised to exercise majority rule. As a consequence, referenda have been used before to determine the provision of public goods (Green et al., 1998; Mitchell & Carson, 2013). We, therefore, proposed in this research to utilize a RCE approach, focused explicitly on whether NSW residents' willingness to pay for management that prevents beach loss.

## 3.2 EXPERIMENT MATERIALS

Following a literature review and two focus group discussions, six attributes and their appropriate levels were identified to characterise beach erosion control policies: beach type, beach distance from residences and beach length and width deterioration (as a percentage), levy time horizon, and annual levy specific to beach category. The payment was presented as a household levy that would apply to all NSW households. This is a familiar payment vehicle for Australians, and in the introduction to the task participants were reminded of the Emergency Services levy (REF) and were informed that the proposed levy would be applied to property or passed along in the form of increased rental payment (the latter made explicit with the intent of informing renters that they would indirectly be affected). The levy would be imposed for a specific time duration, ranging from 10-50 years. The literature indicates that one-off payments can be excessively conservative, which led to our use of the annual levy (MacDonald, Ardeshiri, Rose, Russell, & Connell, 2015; Whitehead & Blomquist, 2006). To arrive at a reasonable range of levies to test, we used an estimate of the net present value of housing at risk under planned beach retreat, the total number of affected households and a 3% annual interest rate over 50 years, say, to calculate upper and lower levy amounts for each beach category. Figure 3 provides an overall upper and lower range of levies, sufficiently wide range to allow coverage for a comprehensive set of future analyses, in terms of population affected by the levy. Table 2 presents the full list of attributes and levels considered for the referendum task.

**Table 1:** Annualised Initial payment for each beach category

| Beach Type | Lower range | Upper range |
|---|---|---|
| **Iconic** | $100 | $660 |
| **Surf** | $20 | $125 |
| **Bay** | $5 | $50 |
| **Main** | $5 | $600 |

**Figure 3:** Wide range of levies used for estimating willingness to pay for erosion control

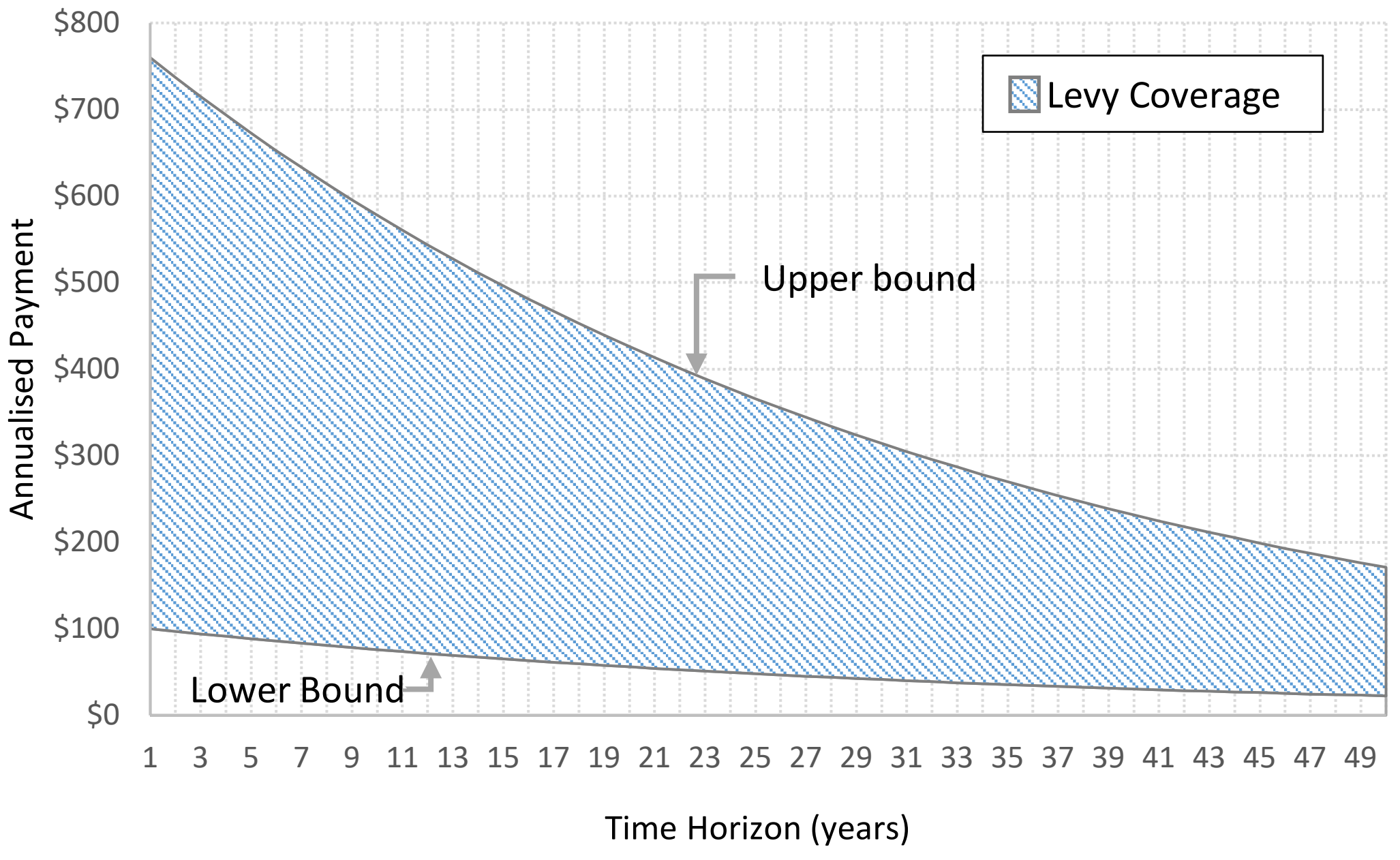

**Table 2:** Attributes and levels used in the referendum task

| Attributes | | levels |
|---|---|---|
| Beach Type | | Iconic beach, Surf beach, sheltered/Bay beach, Main beach |
| Distance from residential location | | 100m, 1km, 10km, 25km, 50km, 100 km, 150km, 200km |
| Time Horizon | | 10 years, 20 years, 40 years, 50 years |
| Sand beach length deterioration | | 5%, 10%, 15%, 20%, 25%, 50%, 75%, 100% |
| Sand beach width deterioration | | 5%, 10%, 15%, 20%, 25%, 50%, 75%, 100% |
| Annual levy to your household ($) | ***Iconic*** | $100, $180, $260, $340, $420, $500, $580, $660 |
| | ***Surf*** | $20, $35, $50, $65, $80, $95, $110, $125 |
| | ***Bay*** | $5, $10, $20, $25, $30, $35, $40, $50 |
| | ***Main*** | $5, $90, $175, $260, $345, $430, $515, $600 |

## 3.3 EXPERIMENTAL DESIGN

Individual policy preferences were measured using the choice modelling framework presented in Figure 4. Individuals could select between the status quo (implying explicit beach deterioration and no specific maintenance action to be taken) and a proposed levy to pay for management that prevents beach loss and maintains the current condition (width and length) of the beach. As mentioned earlier, four beach categories (*Iconic*, *Surf*, *Bay* and *Main)* were studied. Respondents were asked to choose between two options (see Figure 5):

> **Yes**: for a given beach of a certain nominated type and specific proximity to their residential location, an annual levy of the amount shown and for the time horizon specified would be used to counter beach erosion of that beach by preserving the (percentage) length and width indicated.
>
> **No**: this 'status quo' alternative meant that the beach received no specific incremental maintenance action taken by the local council. The consequence of voting for this policy was that the specific beach would suffer a loss of beach length and width over the time horizon ("no specific maintenance action taken" policy scenario on the left in Figure 5). Residents of NSW would not pay any extra levy in this case.

**Figure 4:** Referendum task structure

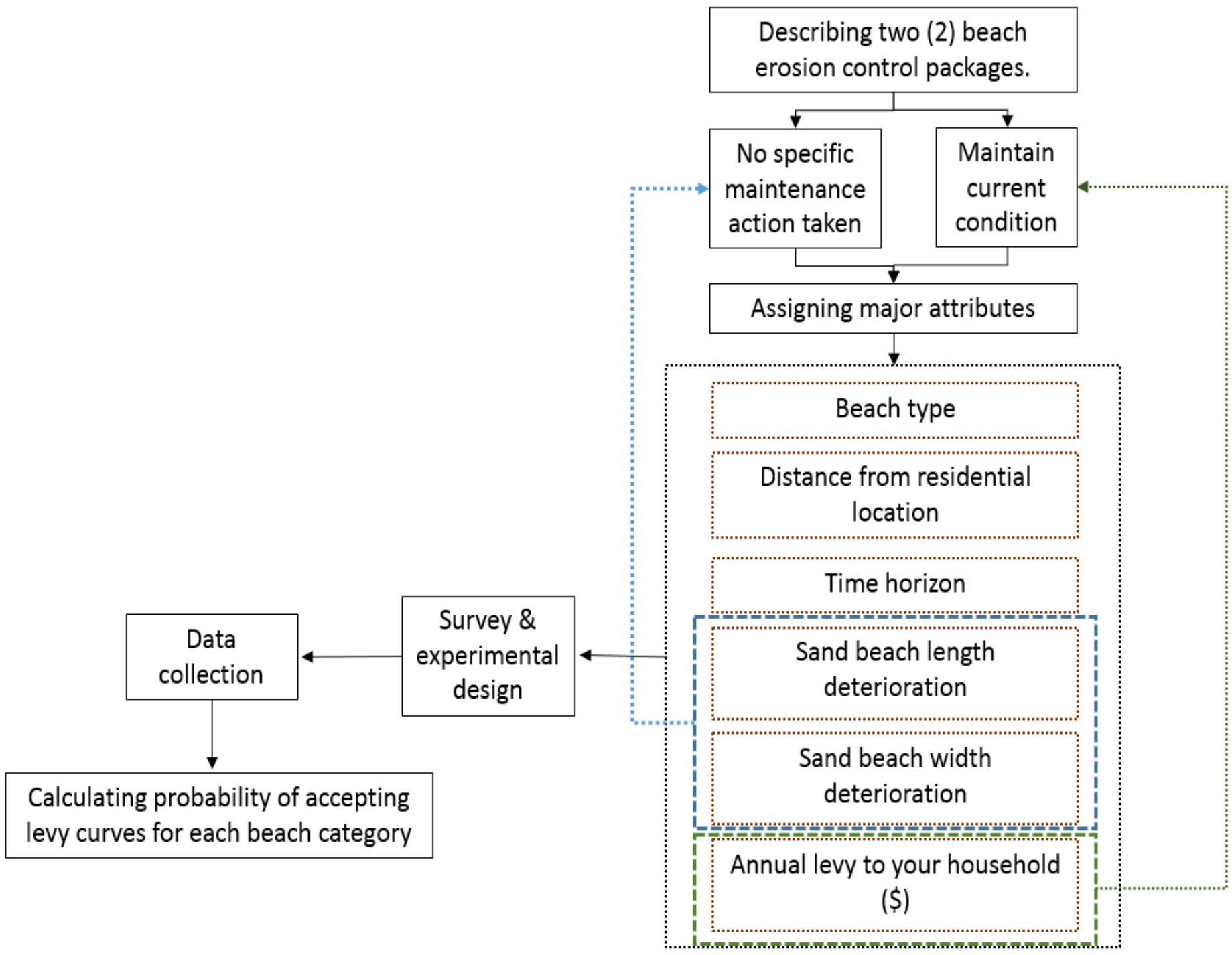

The attribute level values used in specific choice tasks were defined by an efficient experimental design generated using NGENE software. We generated a design using the D-efficiency criteria (Scarpa & Rose, 2008). D-efficiency design strategies produce significantly improved results, in a statistical sense of relative efficiency, then the more traditional orthogonal design (Rose, Bliemer, Hensher, & Collins, 2008).

The final design had a D-error of 0.0185 and included 96 choice tasks in 12 blocks, providing each participant with 8 repeated choice scenarios. Each individual was given 8 hypothetical referenda to complete and was urged to treat each referenda independently of the others. Participants were also reminded to keep in mind their available household income and all other things that this income is spent on. To ensure that the participants took the survey seriously, a short cheap talk script was developed using guidance from Morrison and Brown (2009). Cheap talk is a technique used in SP surveys to remind participants that they should make choices as if they really had to pay. Cheap talk has been shown to be effective at reducing the potential for hypothetical bias in choice experiments (List, Sinha, & Taylor, 2006; MacDonald, et al., 2015; Tonsor & Shupp, 2011). Figure 5 presents an example of the referendum task.

**Figure 5:** Screen capture of an example of the referendum task

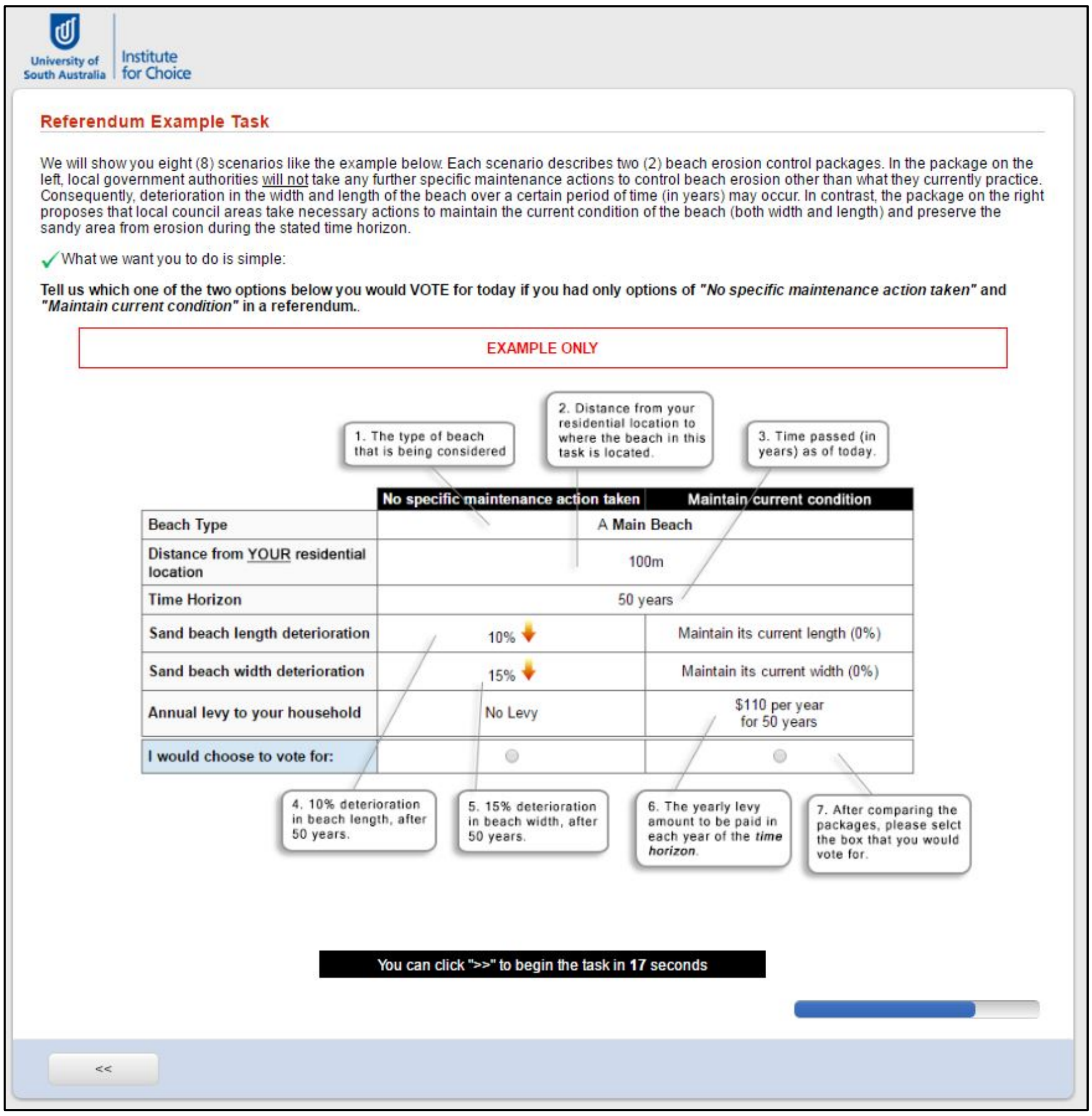

## 3.4 DATA COLLECTION

Data for our analysis came from a state-wide sample of NSW residents. In all, 2014 respondents were drawn from a consumer sample of a major national online panel company. The survey was administered in the period of August 12-22, 2016, through a web-based interface. Respondents were recruited roughly in proportion to the composition of the NSW population in terms of key demographic variables, such as age, gender and income. Figure 6 provides an overview of sample distribution over NSW, with respondents positioned at their postal code centroid.

**Figure 6:** Sample distribution over NSW shown at the postal code centroid.

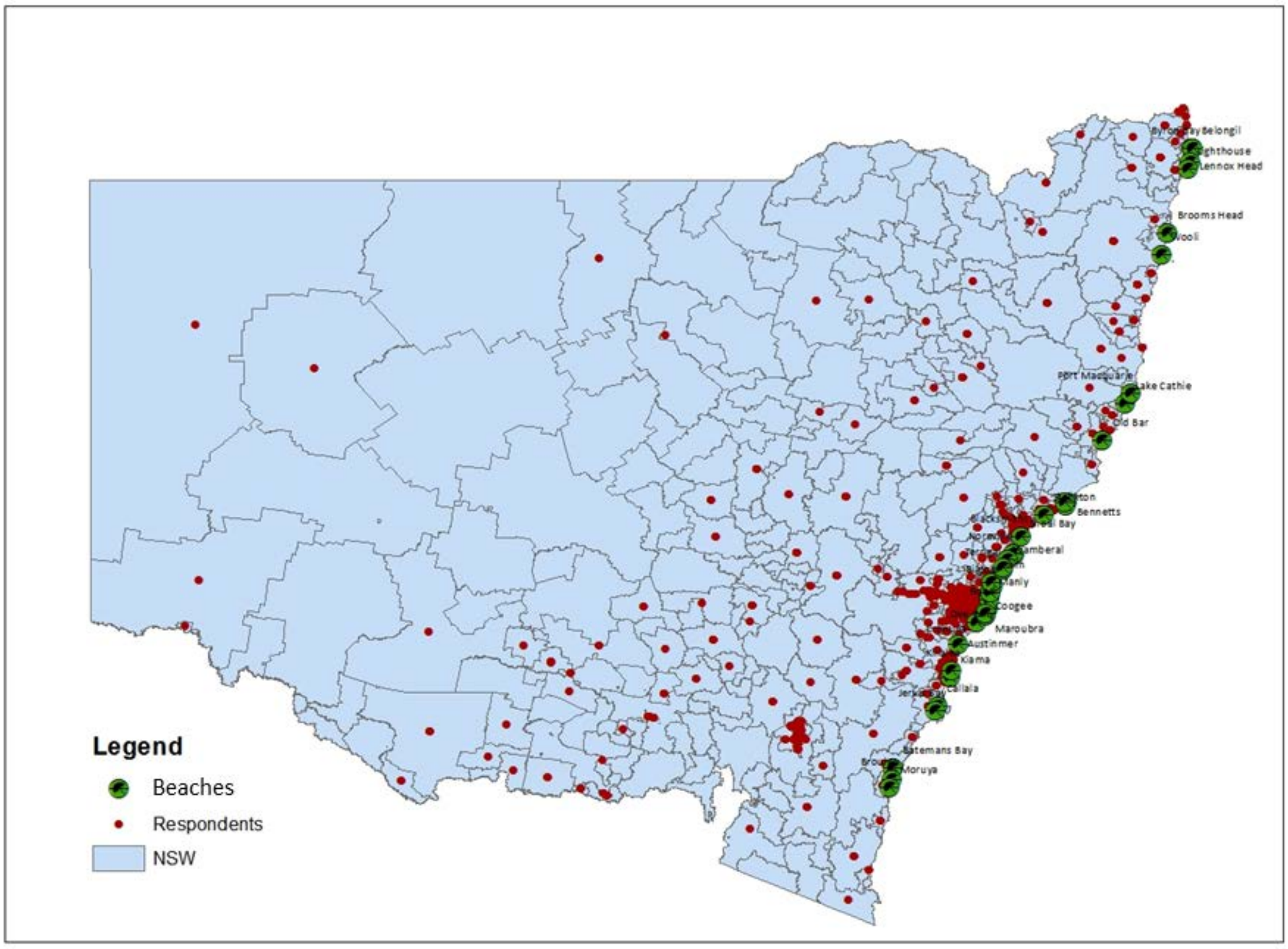

Sample characteristics are given in Table 3. There is a higher female participation (55.6%) compared to male (44.4%). The average age was 49 years. Participants were from different types of households with the majority (34.6%) being "Couple family with children". Among quintiles of income, the largest proportion of respondents (31.4%) falls in the top quintile. Of the 2,014 respondents, 578 (28.7%) indicated that they have not visited a beach in the past 12 months, nor are willing to visit a beach; in this study, such respondents are classified as not being beach users. The remaining 1,436 (71.3%) are considered beach users. This latter group reported that in aggregate they have made 20,007 visits to a pre-specified set of 39 nominated beaches along the NSW coastline (see Figure 66), resulting in an average of 14 visits per year. Almost half of the total visits were made to Iconic beaches, whereas Main (22.5%), Surf (16.5%) and Bay (11.5%) beaches generate the other half of the total visitation. Only 40.7% respondents indicated that they have a degree from a university and 34.4% have an associate degree. Homeowners constitute 68.9% of the sample, and renters the remainder.

**Table 3:** Descriptive statistics for respondents

| Variable | | Statistics |
|---|---|---|
| *Total Participants* | | *2014* |
| *Gender* | | |
| | Male | 44.4% |
| | Female | 55.6% |
| *Age* | | |
| | *18-24 years* | 3.9% |
| | *25-34 years* | 16.5% |
| | *35-44 years* | 21.0% |
| | *45-54 years* | 20.9% |
| | *55-64 years* | 19.9% |
| | *65-74 years* | 15.2% |
| | *75 years and over* | 2.6% |
| *Household type* | | |
| | Couple family with no children | 31.3% |
| | Couple family with children | 34.6% |
| | One parent family | 5.3% |
| | Single person household | 20.5% |
| | Group household | 5.0% |
| | Other Family | 3.3% |
| *Education* | | |
| | Graduate degree | 16.6% |
| | Bachelor's degree | 24.1% |
| | Associate's degree | 34.4% |
| | College graduate or less | 24.9% |
| *Household annual income* | | |
| | Lowest quintile ($1-$33,800) | 6.5% |
| | Second quintile ($33,801-$47,580) | 13.3% |
| | Third quintile ($47,581-$62,190) | 12.5% |
| | Fourth quintile ($62,191-$122,520) | 31.4% |
| | Highest quintile ($122,521 and more) | 26.2% |
| | Prefer not to answer | 10.1% |
| *Dwelling type* | | |
| | Free standing house | 66.5% |
| | Semi-detached, in a row of terrace houses, townhouse | 10.8% |
| | Flat, unit or apartment | 22.0% |
| | Other dwelling (e.g. caravan, cabin, houseboat, or improvised home) | 0.7% |
| *Is this dwelling…?* | | |
| | Owned | 68.9% |
| | Rented | 31.1% |
| *Employment* | | |
| | Full-time | 38.7% |
| | Part-time | 19.3% |
| | Retired | 22.5% |
| | Un-employed | 4.7% |
| | Not in labour force | 14.8% |

# 4. DATA ANALYSIS

A concern with hypothetical referendum tasks is the possibility of strategic or protest voting in the form of “Nay-saying” and “Yea-saying” (i.e., voting ‘no’ irrespective of policy attributes variation, and voting ‘yes’ no matter the policy attributes). Among psychologists and sociologists studying response acquiescence, yea-saying is defined as the tendency to agree with questions regardless of content. Visa Versa the tendency to disagree is distinct as Nay-saying (Blamey, Bennett, & Morrison, 1999; Couch & Keniston, 1960; Moum, 1988). Traditional statistical analyses of DCEs do not handle these extreme preferences well. Recognising this limitation, the random utility choice model utilised for this study is based on an innovative use of a standard Latent Class (LC) model.
To begin, the proposed model allows the sample to be separated between those who make trade-offs and those who don’t; among those who don’t make trade-offs, it makes a distinction between those unswervingly protesting against or in favour of the referenda. For those who make trade-offs, it is assumed that the individual may be decomposed into discrete segments that differ in their predisposition towards beach maintenance policy and their sensitivity to different attributes presenting the policy. Thus, in addition to handling trade-off heterogeneity among “traders”, we allow one segment to represent the Yea-sayers and another segment to represent the nay-sayers.

Figure 7, illustrates a path diagram of the underlying structural model representing the choice process. Sociodemographic characteristics and individual’s choice behaviour in response to a given set of policy are the observable – or manifest - variables (presented in rectangular shapes). Following Swait (1994) we allow sociodemographic characteristics to form the “segment membership” as well as “taste in preference” in residents’ choices. Structural latent variables are depicted through ellipses.

(1) S*ociodemographic characteristics* form the latent segment membership likelihood functions for an individual.
(2) Through a latent *segment classification mechanism,* the membership likelihood functions determine the latent segment (i.e. yea-sayers, nay-sayers and traders) to which an individual belongs.
(3) The decision-maker has preferences with respect to the policy which determine the yes/no vote. These preferences are determined by the individual’s perceptions of the given attributes, his/her personal characteristics and the latent class to which he/she belongs. These preferences are conditional on, and specific to, the segment to which the person belongs.

This structural model is an adaptation of the general framework presented in McFadden (1986) and Swait (1994). The latent class model (LCM) has been used extensively for the analysis of individual heterogeneity (for theoretical discussion see Boxall & Adamowicz, 2002; Greene & Hensher, 2003; Heckman & Singer, 1984; Swait & Adamowicz, 2001).

**Figure 7:** A structural equation model of latent segmentation and choice process. Partially adapted from Figure 1, Swait (1994).

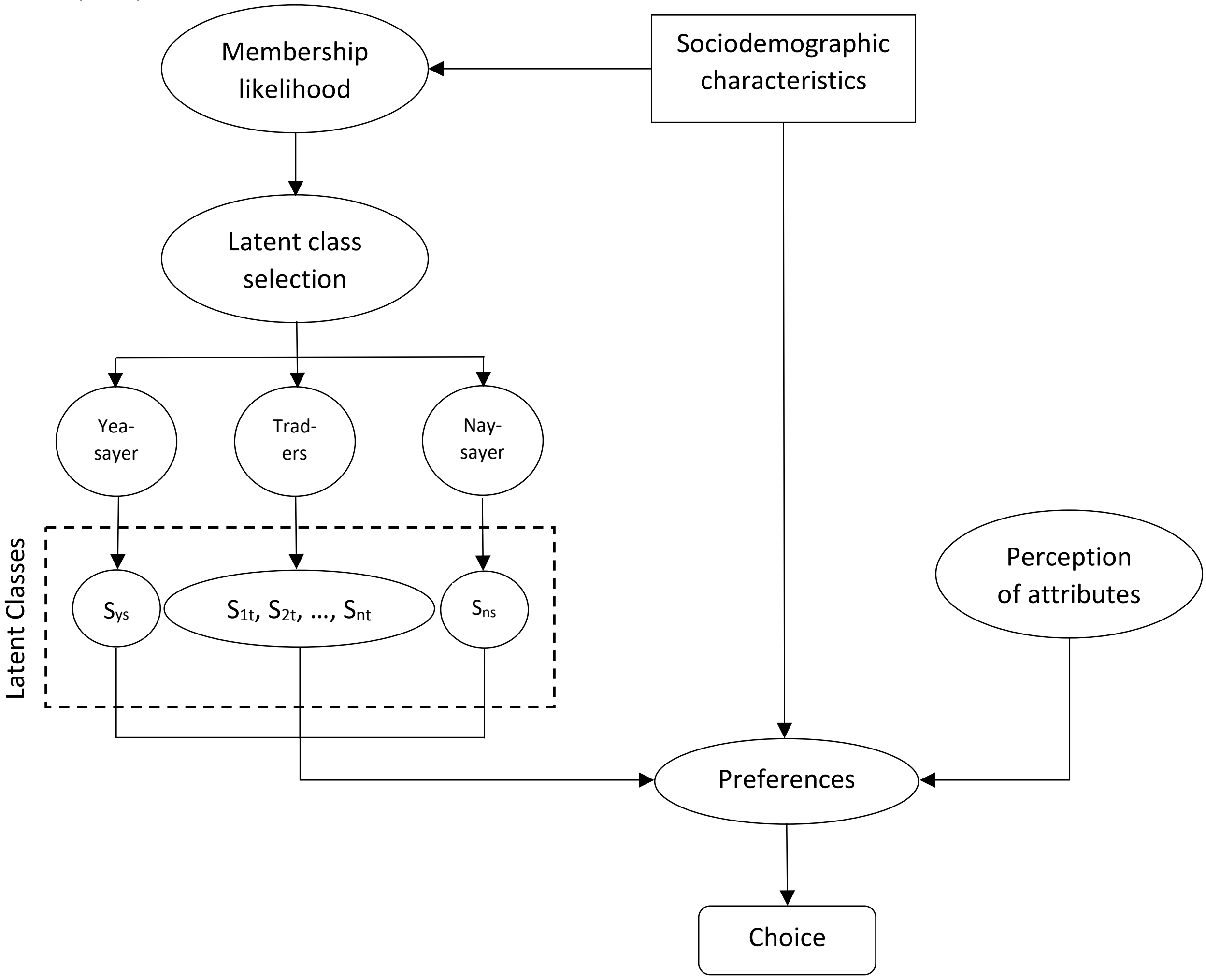

## 4.1 Model formulation

The underlying theory of the LCM posits that individual behaviour depends on observable attributes and on latent heterogeneity that varies with factors that are unobserved by the analyst. We propose to analyse this heterogeneity through a model of discrete parameter variation. Thus, it is assumed that individuals are implicitly sorted into a set of *S* classes (whether known or not to that individual), but which class contains any particular individual is unknown to the analyst. Having said that, by definition, we expect from the yea-sayers - who always agree to pay a levy to maintain the beach regardless of the costs and benefits – to be deterministic and have probability value equal to one for the Yes alternative. For this reason, we allow the utility for paying a levy ($U_{ivs}$ to be a very large positive value (effectively, positive infinity) for this segment. Visa Versa, by fixing $U_{ivs}$ to a large negative value (effectively, negative infinity) we force the response of individuals in this segment to be No with probability one; conversely, the nay-saying segment has zero probability of voting Yes. The segment(s) who make trade-offs between the given options are considered to have a finite (and to be estimated) $U_{ivs}$ as specified in equation (1).

$$U^*_{i|s} = \begin{cases} \text{If } \textit{Yea-sayers} & U^*_{i|s} = +\infty \\ \text{If } \textit{Trader} & -\infty < U^*_{i|s} < +\infty \\ \text{If } \textit{Nay-sayers} & U^*_{i|s} = -\infty \end{cases} \quad (1)$$

To establish certain components required to build the model, we assume that a choice scenario presents alternatives in choice set $C_r$, r=1,…,R, where R is the number of choice scenarios in a choice experiment. Each alternative *i* has utility

$$U_{ir|s} = V_{ir|s} + \varepsilon_{ir|s}, i \in C_r, \quad (2)$$

where $V_{ir|s}$ is the systematic utility for the alternative in the r[th] scenario, conditional on belonging to class *s* (=1,…,S) with a set of preference component $\beta_s$ and an individual's sociodemographic characteristics $Z_i$ such that

$$V_{ir|s} = \beta_s X_{ir|s} + Z_i \sigma_{ir|s}, \quad (3)$$

and $\varepsilon_{ir|s}$ is the stochastic utility of the alternative. If we assume that the $\varepsilon_{ir|s}$ is independently and identically Gumbel distributed with scale $\mu$, the class conditional choice model is a multinomial logit formulation:

$$P_{ir|s} = \frac{\exp(\mu V_{ir|s})}{\sum_{j \in C_r} \exp(\mu V_{jr|s})}. \quad (4)$$

We assume that given the class assignment, the $R_i$ events are independent. This is possibly a strong assumption, especially given the nature of the sampling design used in our application—a stated choice experiment in which the individual answers in sequence, and in short order, repeated choice scenarios. In fact, there might well be correlation in the unobserved parts of the random utilities. The latent class does not readily extend to deal with this potential autocorrelation, so we have left this aspect for further research. Thus, for the given class assignment, the contribution of individual *i* to the likelihood would be the joint probability of the sequence $P_i$=[ $P_{i1}$, $P_{i2}$,..., $P_{ir}$]. This is

$$P_{i|s} = \prod_{r=1}^{R} P_{ir|s} \tag{5}$$

The class assignment is unknown. Let $H_{is}$ denote the prior probability for class *s* for individual *i*. The polytomous multinomial logit form is

$$H_{is} = \exp(z_j'\theta_s) / \textstyle\sum_{s=1}^{S} \exp(z_j'\theta_s) \ , s = 1, \ldots, S, \theta_S = 0, \tag{6}$$

where $z_i$ denotes a set of observable characteristics which enter the model for class membership. Note that not all θ's can be identified since the corresponding variables do not vary from class to class. Hence, one must normalize one of these vectors to a constant, say, zero. We have chosen to normalize the θ for the last class, S. The likelihood for individual *i* is the expectation (over classes) of the class-specific contributions:

$$P_i = \sum_{s=1}^{S} H_{is} P_{i|s} \tag{7}$$

The Log likelihood for the sample is

$$lnL = \textstyle\sum_{i=1}^{N} \ln P_i = \sum_{i=1}^{N} ln \left( \sum_{s=1}^{S} H_{is} \prod_{r=1}^{R} P_{ir|s} \right) \tag{8}$$

Maximization of the log likelihood with respect to the *S* structural parameter vectors, $\beta_s$ and the *S*−1 latent class parameter vectors, $\theta_s$ is a conventional problem in maximum likelihood estimation.

# 5. Results

## 5.1 Understanding preferences of different survey segments

A latent class model was used to estimate individual policy preferences for beach maintenance. Results have been weighted by age and gender to represent the NSW population.

The latent class choice model used to represent the probability that respondents are willing to pay a levy identified four segments in the population. Segment one was "Nay-sayers", who were unwilling to pay any levy at all, no matter the amount or beach type; the second segment was the "Yea-sayers", who said *yes* to any amount of levy payment. Segments 3 and 4 constitute the trade-off segments that displayed preference heterogeneity around paying a levy for different beach categories over a time horizon. Based on the estimated results and to make it easier for the readers to follow the estimation results, hereinafter we call Segment 3 the ***nay-leaning*** and segment 4 the ***yea-leaning groups.*** Reasons for this naming will become more obvious after reading the following sections, which explain and compare estimates of willingness to pay in response to different attributes (levies of different amounts and for different durations) amongst the segments. Segment membership propensities are also calculated and reported to allow policymakers to understand the relative distribution of the four respondent segments within the broader population. On average, the model predicts that 5% of the population belong to the yea-sayers segment, 25.5% belong to the yea-leaning segment, 33.9% belong to the nay-leaning segment, and 35.6% belong to the Nay-sayer segment.

**Figure 8:** Population segments and associated membership probabilities

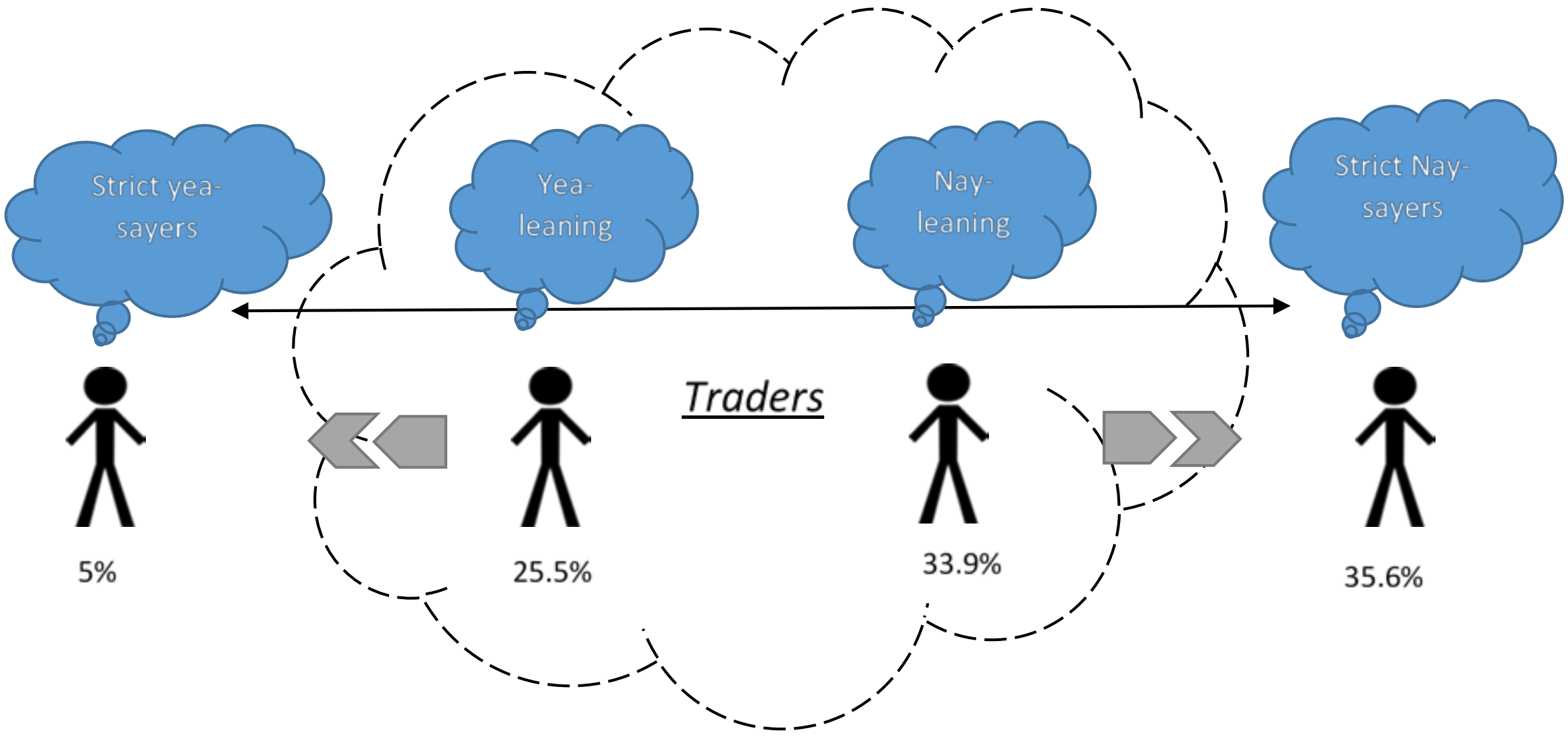

Table 4 provides the parameter estimation results for the latent class model. Linear and non-linear (quadratic) transformation of the continuous variables were both expressed in the utility functions. Further, to avoid collinearity of the linear and quadratic terms, *orthogonal polynomial* coding was used (for more information refer to chapter 9 in Louviere, Hensher, & Swait, 2000).

Based on the estimations and results in Table 4, respondents belonging to the ***nay-leaning*** segment are more reticent about paying levies: they are willing to pay a levy only in a reduced and specific set of circumstances, and, therefore, a range of different attributes appear important in their utility function. In contrast, respondents belonging to the ***yea-leaning*** segment display a utility that is only affected by the levy amount. This indicates that they are willing to pay some acceptable amount of levy, regardless of their distance to the beach, the payment period, or beach length and width deterioration. Surprisingly, beach length deterioration was only significant for Bay beaches and respondents belonging to the ***nay-leaning*** group (Figure 14), and width deterioration was not significant for any beach category. This suggests that the respondents were willing to pay some levy amount to maintain a beach *in its current state*, irrespective of the expected severity of erosion impacts on sand volumes – even to avoid a 5% loss of sand (minimum attribute level, Table 2).

**Table 4:** Parameter estimates for the referendum task

| | Parameters | Value | Std err | t-test | p-value |
|---|---|---|---|---|---|
| **Yea-sayers** | | | | | |
| | Constant | **Fixed (at +∞)** | | | |
| **Nay-sayers** | | | | | |
| | Constant | **Fixed (at -∞)** | | | |
| ***Nay-leaning*** | | | | | |
| | Constant | -2.02 | 0.55 | -3.67 | <0.001 |
| *Iconic beach* | | | | | |
| | Constant | 0.0284 | 0.521 | 0.05 | |
| | Levy | -0.456 | 0.129 | -3.53 | <0.001 |
| | Time horizon | -0.507 | 0.0652 | -7.77 | <0.001 |
| | Distance to beach | -0.319 | 0.133 | -2.4 | 0.02 |
| *Surf beach* | | | | | |
| | Constant | 2.18 | 0.541 | 4.03 | <0.001 |
| | Levy | -0.538 | 0.0597 | -9.01 | <0.001 |
| | Time horizon | -0.122 | 0.0405 | -3.01 | <0.001 |
| | Distance to beach | -0.136 | 0.0517 | -2.63 | 0.01 |
| *Bay beach* | | | | | |
| | Constant | 0.737 | 0.559 | 1.32 | 0.19 |
| | Length deterioration | 0.455 | 0.149 | 3.05 | <0.001 |
| | Levy | -1.01 | 0.0881 | -11.41 | <0.001 |
| | Time horizon (quadratic form) | -0.74 | 0.203 | -3.64 | <0.001 |
| | Distance to beach | -0.534 | 0.124 | -4.31 | <0.001 |
| *Main beach* | | | | | |
| | Constant | **Fixed (at zero)** | | | |
| | Levy | -0.525 | 0.266 | -1.98 | 0.05 |
| | Distance to beach | -0.826 | 0.442 | -1.87 | 0.06 |
| ***Yea-leaning*** | | | | | |
| | Constant | 1.08 | 0.128 | 8.4 | <0.001 |
| *Iconic beach* | | | | | |
| | Constant | -1.52 | 0.139 | -10.98 | <0.001 |
| | Levy | -0.742 | 0.0622 | -11.93 | <0.001 |
| *Surf beach* | | | | | |
| | Constant | -0.376 | 0.166 | -2.27 | 0.02 |
| | Levy (quadratic form) | -1.18 | 0.128 | -9.23 | <0.001 |
| *Bay beach* | | | | | |
| | Constant | 0.33 | 0.161 | 2.05 | 0.04 |
| | Levy | -0.558 | 0.0749 | -7.45 | <0.001 |
| *Main beach* | | | | | |
| | Constant | **Fixed (at zero)** | | | |
| | Levy | -0.699 | 0.0634 | -11.02 | <0.001 |
| **Segment membership** | | | | | |
| *Yea-sayers* | | | | | |
| | Constant | -1.62 | 0.125 | -12.95 | <0.001 |
| *Nay-sayers* | | | | | |
| | Constant | 0.254 | 0.0902 | 2.82 | <0.001 |
| | Female gender | -0.167 | 0.0521 | -3.2 | <0.001 |
| | Beach-users | -0.286 | 0.0548 | -5.21 | <0.001 |
| | Age | 0.185 | 0.0528 | 3.5 | <0.001 |
| *Nay-leaning* | | | | | |
| | Constant | 0.285 | 0.111 | 2.57 | 0.01 |
| *Yea-leaning* | | | | | |
| | Constant | **Fixed (at zero)** | | | |
| **Estimation report** | | | | | |
| | Number of estimated parameters | **30** | | | |
| | Sample size | **16112** | | | |
| | Log likelihood | **-6839.15** | | | |

103

### 5.1.1 WILLINGNESS TO PAY FOR DIFFERENT BEACH TYPES

Willingness to pay (WTP) for management that prevents beach loss differed by segment and by beach type (Table 5). Overall WTP was highest for 'Iconic' and 'Main' beach types. The WTP figures presented in Table 5 represent state-wide averages, whereby the proportion of households in the state that would be considered yea-sayers, yea-leaners, nay-leaners and nay-sayers (based on their demographic characteristics) has been accounted for in the calculation of the average WTP per household. In practice, when calculating WTP for a specific beach in a specific location, the relative proportion of households belonging to each population segment would be calculated with reference to local socio-demographic characteristics. This means that the overall WTP and preference for different beach types may vary across different parts of the state.

**Table 5:** Average household willingness to pay for management that prevents beach loss, by beach type and population segment

| | **NSW household average WTP** ($ per annum) | **Yea-sayers*** ($ per annum) | **Yea-leaners** ($ per annum) | **Nay-leaners** ($ per annum) | **Nay-sayers** ($ per annum) |
|---|---|---|---|---|---|
| Bay | $15.66 | $50 | $29.05 | $17 | $0 |
| Surf | $31.90 | $125 | $50.07 | $38.0 | $0 |
| Main | $100.58 | $600 | $173.07 | $78.0 | $0 |
| Iconic | $161.40 | $660 | $257.06 | $185.4 | $0 |

* For yea-sayers, WTP is dependent on the upper bound of levy amount associated with each beach type (see Table 2)

### 5.1.2 WILLINGNESS TO PAY A MANAGEMENT LEVY

We also investigated WTP by exploring the proportion of households in each population segment that was willing to pay a levy at different amounts and/or for different lengths of time. This provides additional important information to policy-makers about the degree to which the introduction of a levy for management that prevents beach loss would be considered acceptable amongst different segments of the population.

Figure 9 indicates that the proportion of households that are willing to pay a levy is greatest for Bay beaches, followed by Surf, Iconic, then Main beaches. This trend is the same for respondents in both ***nay-leaning*** & ***yea-leaning*** (Figures 11 & 12) groups; however, the proportion of households that are willing to pay the levy varies between the ***nay-leaning*** & ***yea-leaning*** segments. It is important to note that the lower proportion of households that is willing to pay a levy for the "*<u>all respondent</u>*" scenario (Figure 9) compared to ***nay-leaning*** & ***yea-leaning***, is due to the presence of nay-sayers (which constitute about one-third of respondents). This highlights the problem that our segmentation approach and the structural equation model illustrated in Figure 7 is trying to address.

***Nay-leaning*** and ***yea-leaning*** groups showed different responses to increasing levy amounts at different beach types. Bay beaches were associated with the highest rate of levy payment, with a high of ~95% of households from both segments being willing to pay a levy amount of $5. This reduced at a fairly similar rate to a low of 40% of ***nay-leaning*** and 57% of ***yea-leaning*** households willing to pay the maximum levy amount of $50. For Surf beaches, the ***yea-leaning*** group had a much higher proportion of households that were willing to pay lower levy amounts (peak of 89% compared to 61% WTP a levy of $5 for ***nay-leaners***). This declined to 19% and 15% of households WTP a levy amount of $125 for ***nay-leaning*** and ***yea-leaning*** groups respectively. The greatest difference between these two groups was observed for WTP for Main and Iconic beaches. ***Nay-leaners*** had a relatively low proportion of households willing to pay any levy amount, with 5% to 25% of households willing to pay for management at Iconic beaches and 2%-15% of households willing to pay for management of Main beaches. In contrast, ***yea-leaners*** had much higher levy acceptance rates, with 13% to 74% of households being willing to pay a levy for Iconic beaches, and from 20% to 83% willing to pay a levy for Main beaches.

**Figure 9:** Proportion of households that are willing to pay the different annual levy amounts– *All respondents.*

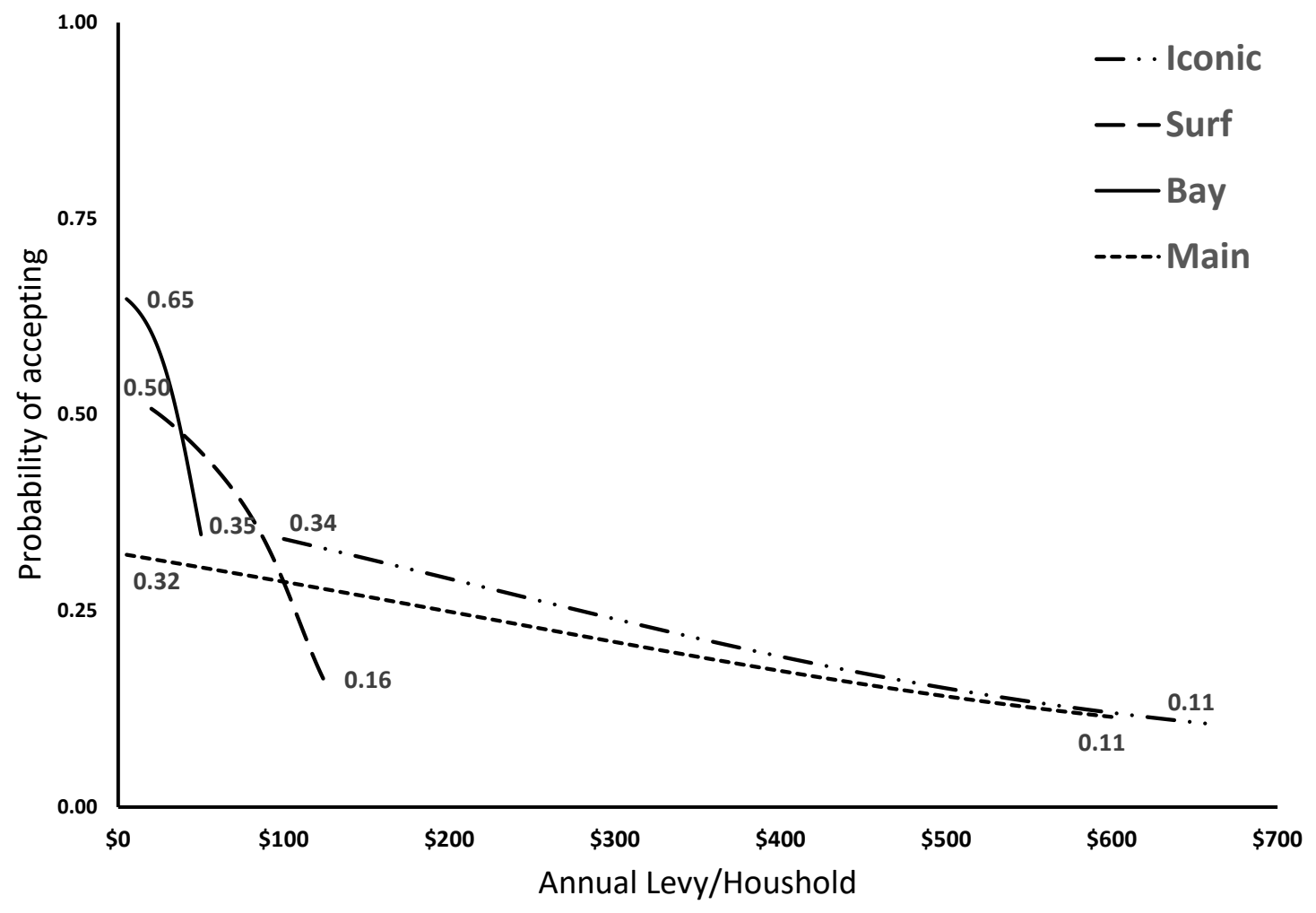

**Figure 10:** Proportion of households that are willing to pay the different annual levy amounts – *nay-leaning.*

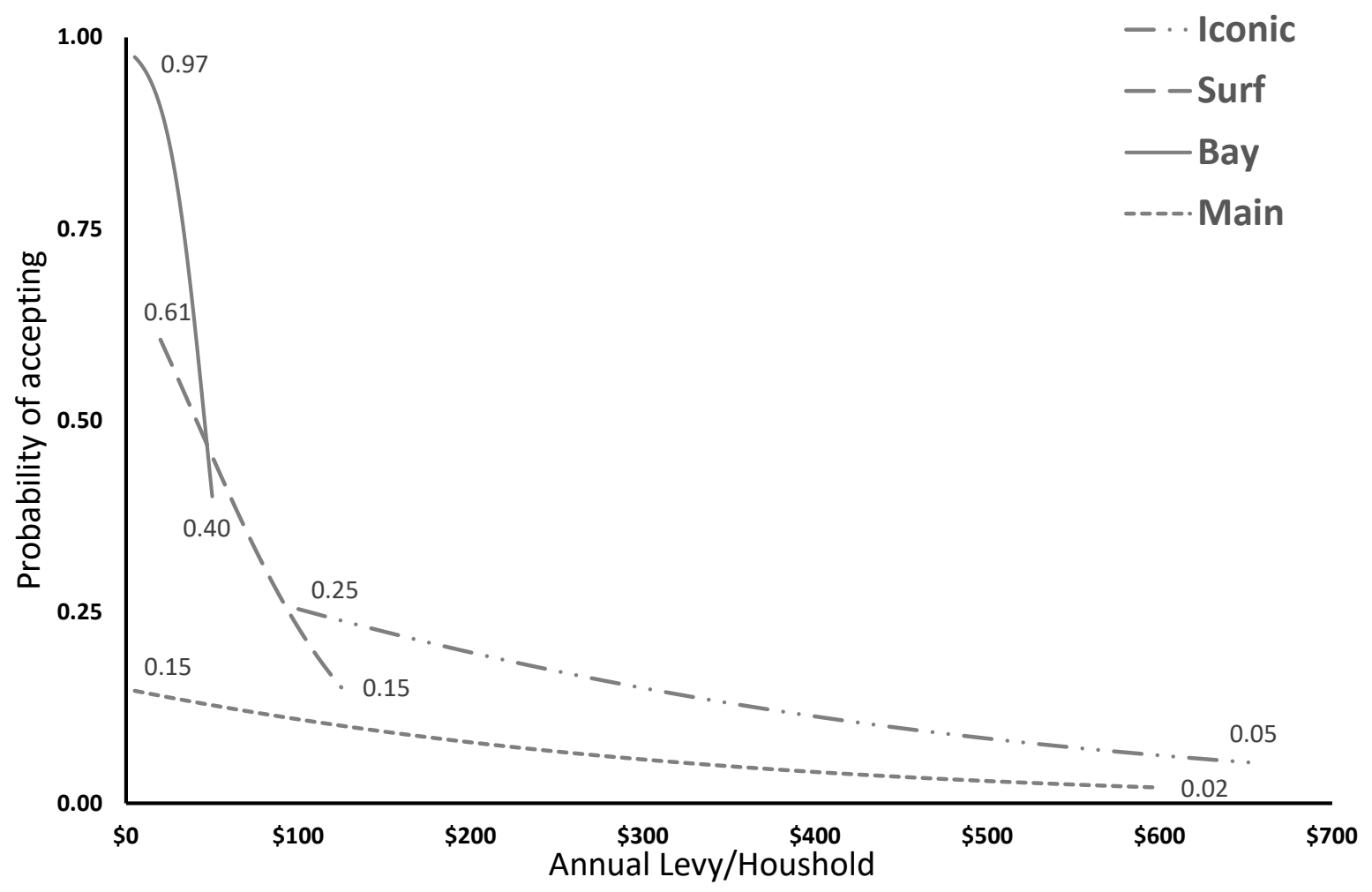

**Figure 11:** Proportion of households that are willing to pay the different annual <u>levy</u> amounts – *yea-leaning*

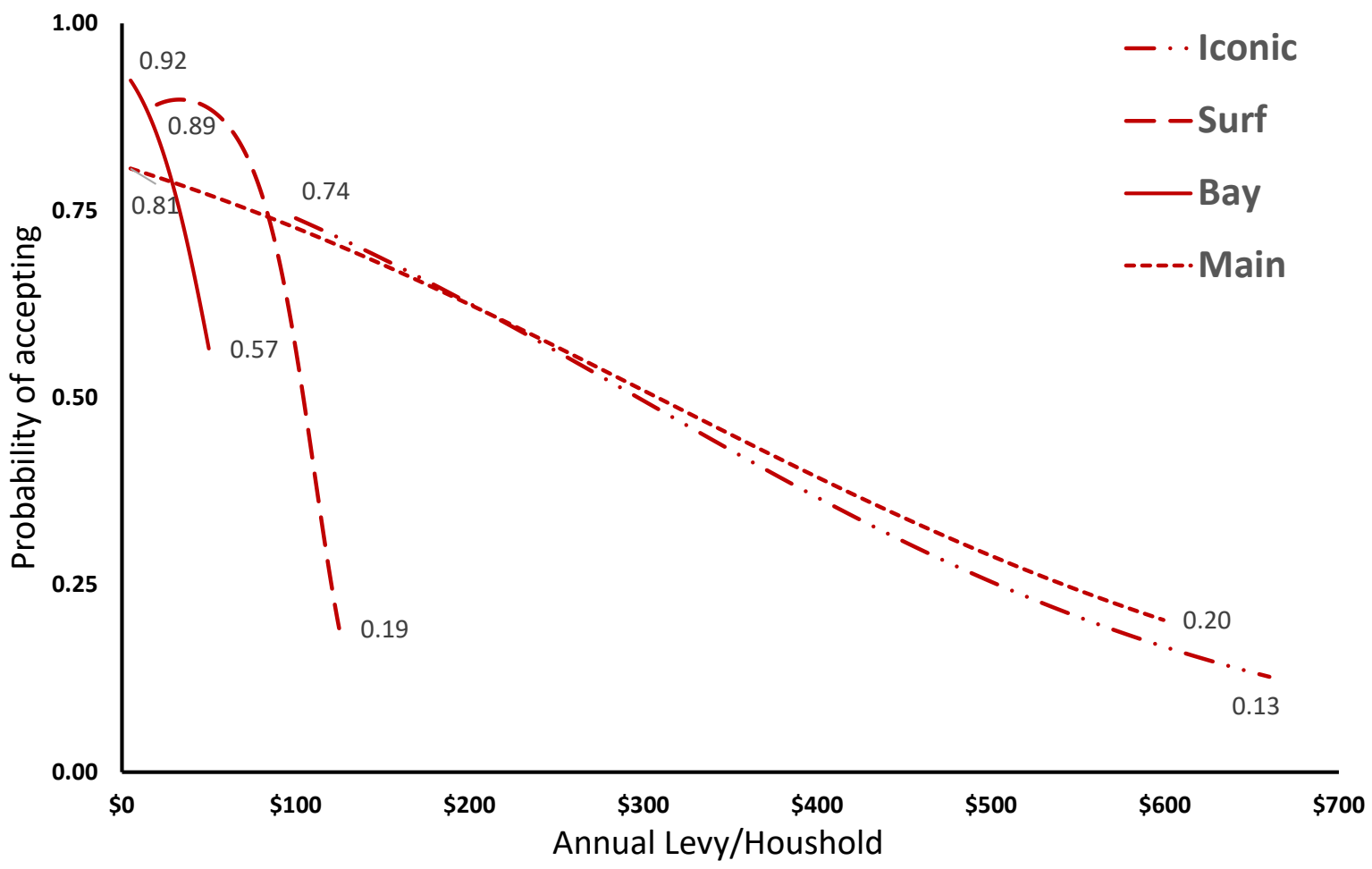

### 5.1.3 Variation in willingness to pay a management levy in response to distance to beach, payment horizon and length deterioration

For the ***yea-leaning*** segment, WTP the management levy was not influenced by distance from the respondents' home to the beach in question, nor the payment horizon and neither the length or width deterioration (Table 4). This suggests that this segment's WTP is not influenced by the likelihood that they will visit a specific site, nor bounded by any specific time period. It follows that non-market values held by the yea-leaning segment are likely to be dominated by 'existence' and 'bequest' value (following Marre et al. 2015; see Discussion).

In contrast, distance to the beach and the payment time horizon had a significant influence on WTP for the ***nay-leaning*** segment. Distance to the beach had a significant effect for all four of the beach categories investigated. As the distance to beach increases, the probability of ***nay-leaning*** households being willing to pay the average levy for each beach category declined. The rate of decline was greatest for Main and Bay beaches, and less for Surf and Iconic beaches (Figure 12). For the ***nay-leaning*** segment, an increasing time horizon had a negative effect on WTP for iconic, surf and bay beaches (Figure 13).

Finally, for the ***nay-leaning*** segment, the beach length deterioration was only significant for Bay beaches (Figure 14), and width deterioration was not significant for any beach category.

Taken in tandem, the trends in WTP in response to the distance from beach and payment horizon for the ***nay-leaning*** segment suggest that these respondents hold different values for different types of beaches. Lower sensitivity to the payment horizon for Bay and Main beaches (and Surf beaches – although to a lesser degree) suggests that values for these beaches may include some portion of bequest value. High sensitivity to the distance from the beach for Bay and Main beaches types suggests ***nay-leaning*** respondents may also hold 'option value' for these two beach types (being more willing to conserve beaches they are more likely to visit). These different non-market values and their implications for policy and management are discussed further in Section 6.

**Figure 12:** Proportion of households that are willing to pay the average levy of each beach category as <u>*distance*</u> to beach from residential location increases – *nay-leaning*

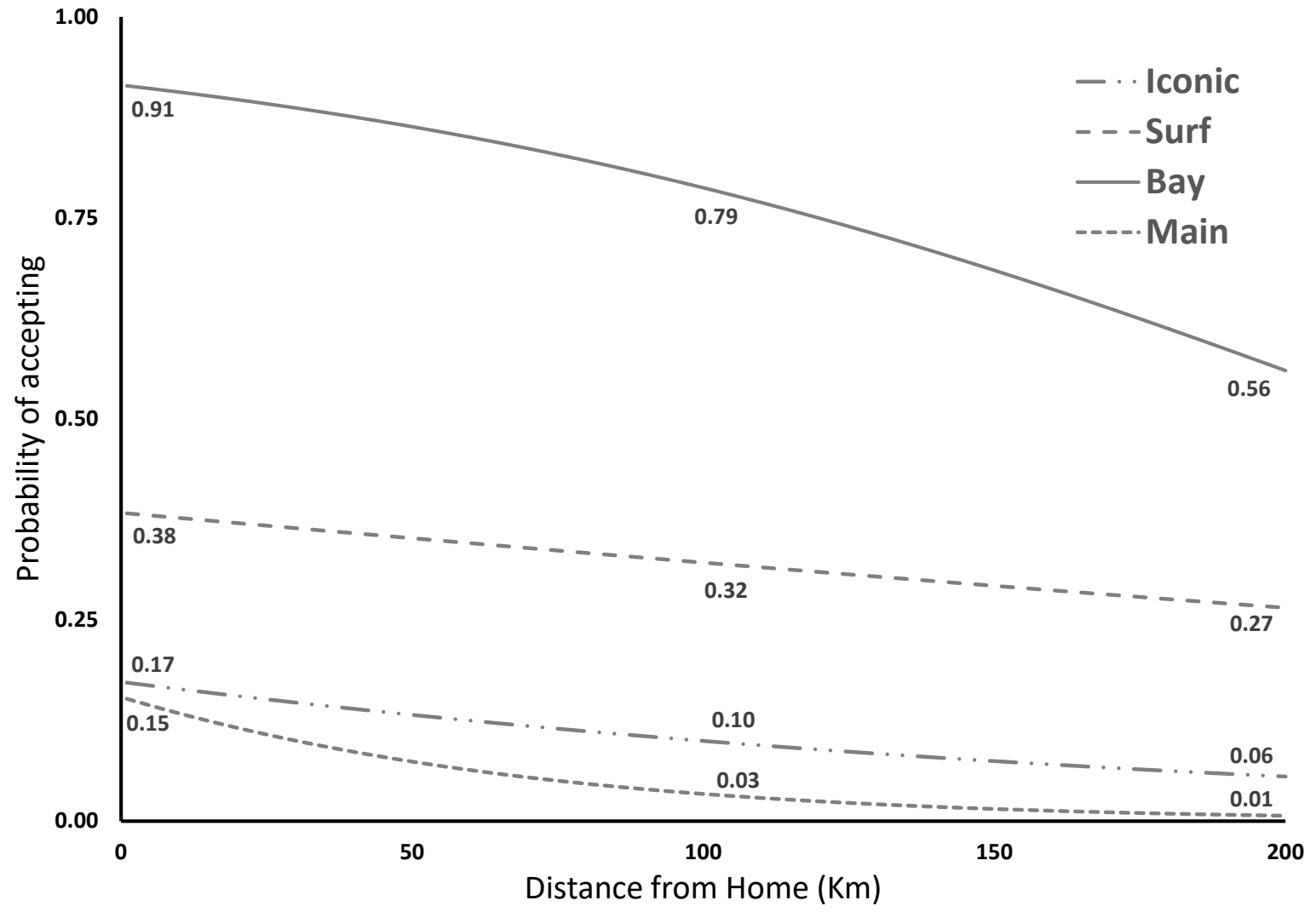

**Figure 13**: Proportion of households that are willing to pay the average levy of each beach category as <u>*time horizon*</u> for paying the levy increases – *nay-leaning*

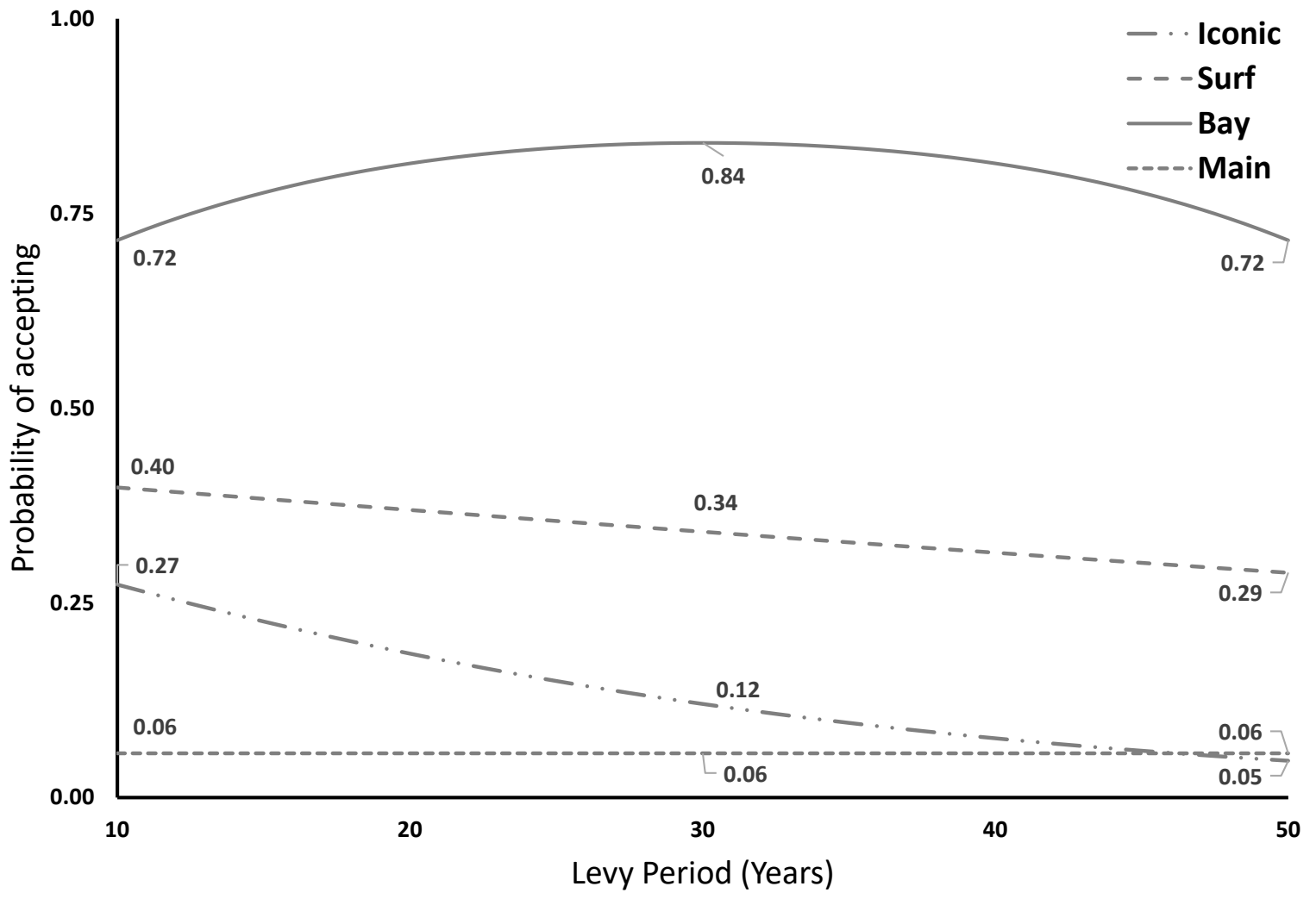

**Figure 14:** Proportion of households that are willing to pay the average levy amount as percentage of <u>*length deterioration*</u> increases – *nay-leaning*

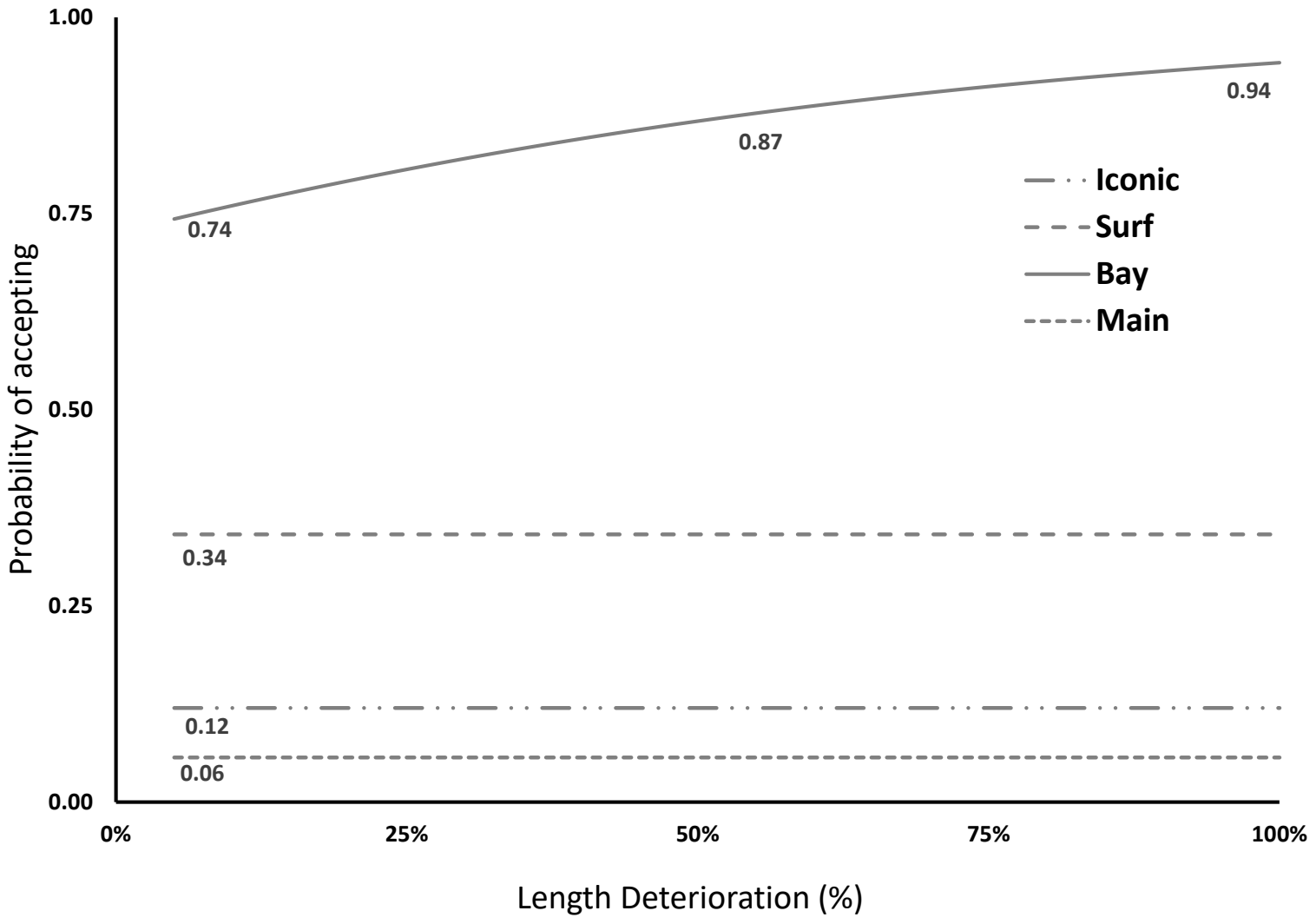

## 5.2 Understanding the survey population

Results in the lower section in Table 4 show segment membership propensity. The estimation results indicate that the probability of membership to different survey segments is not strongly related to respondents' characteristics. The exception is for the nay-sayers group. Characteristics such as gender, being a beach user and increasing age became significant in positioning a respondent to be more likely in the nay-saying group. Males relative to females, non-beach users relative to beach-users and older relative to younger individuals, all increase the probability of saying ***no*** to any kind of levy payments.

After weighting the sample to the NSW population, 35.6% of the population are classified as nay-sayers, and 5% as pure yea-sayers, implying that the remaining 59.3% would trade off policy attributes with different marginal rates of substitution (Figure 15). This 59.3% cluster is constituted of 33.9% ***nay-leaning*** and 25.5% ***yea-leaning*** segments. In Figure 16 we present the probability of belonging to different segments of Male respondents only. As shown the probability of belonging to the Nay-sayers group when all respondents are included (35.6%) increases to 39.3% for male only. Further, if the male is not a beach-user (Figure 16B) this probability increases to 46.0% and if this non beach-user male belongs to 75 and plus age category, the probability of belonging to nay-sayers increases to 55.1% (Figure 16C).

**Figure 15:** Average class membership propensity for the NSW population.

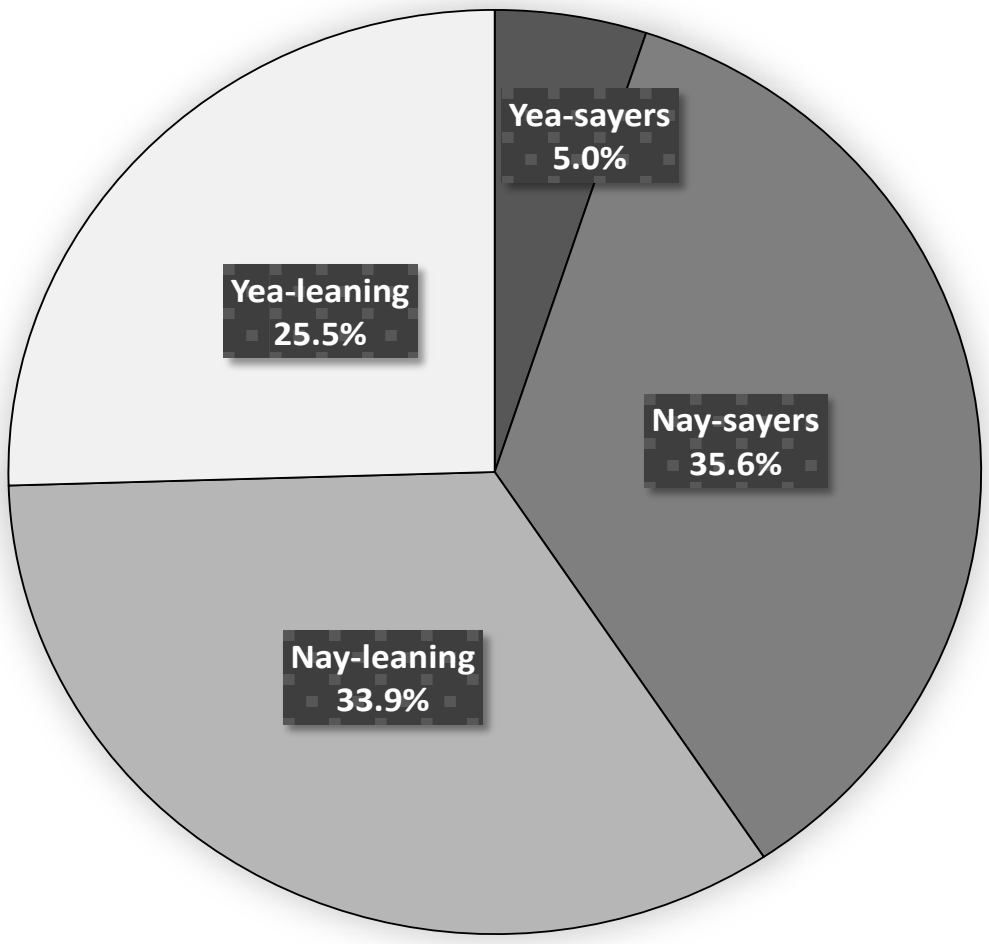

**Figure 16:** Scenario example of the possibility of belonging to the Nay-sayers group by variation in individual characteristics.

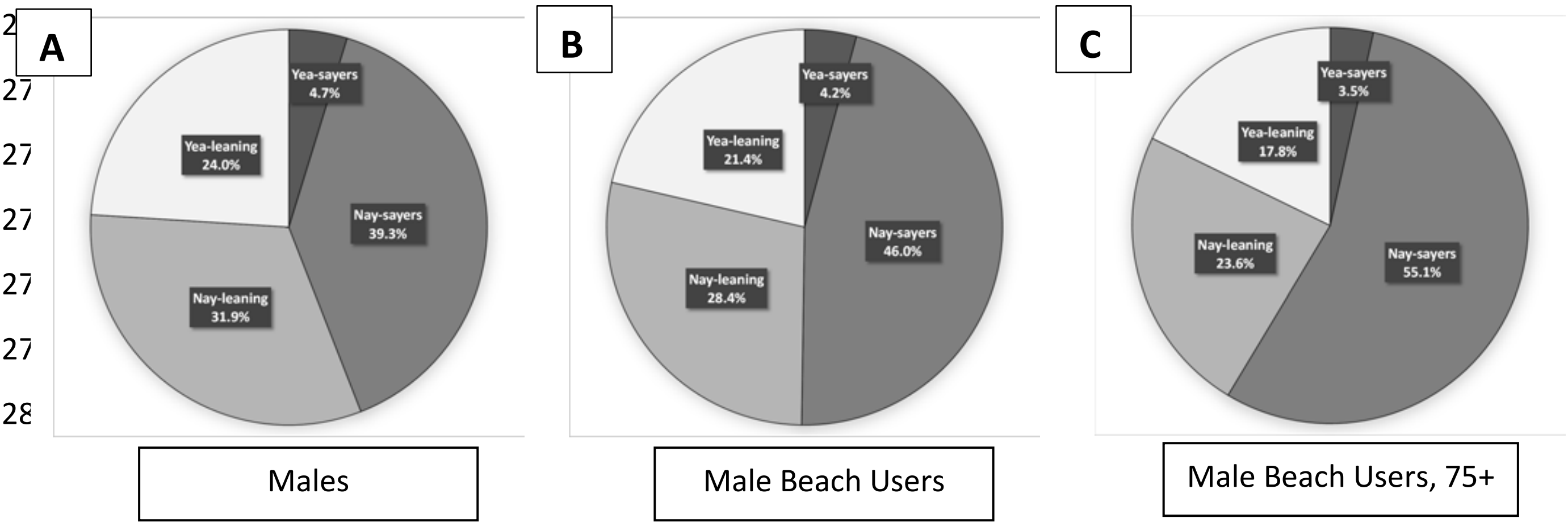

# 6. DISCUSSION

## 6.1 NON-MARKET VALUES FOR NSW BEACHES & IMPLICATIONS FOR MANAGEMENT

Results of the referendum choice experiment presented in this study provide an estimate of WTP for management to preserve sandy beaches in NSW in the face of coastal erosion. We find that 65% of the population would be willing to pay some amount of levy, dependent on the policy setting. Like other CV and CM studies, our study provides an estimate of non-market value. Given that our referendum questioned about respondents' willingness to pay a levy irrespective of beach visitation, we consider that it represents non-use value. Moreover, because it was clear in our survey that any levy would be additional to the travel costs that individual respondents would incur to access a given beach, we consider that the non-use values estimated in this study are additional to use (e.g. recreation) values. The non-use values associated with the preservation of sandy beaches estimated in our study can be used as an input to cost-benefit analysis of coastal management options along the NSW coastline (and elsewhere if benefit transfer is undertaken in an appropriate manner). This would enable the economic outcomes associated with different configurations of built and natural assets to be assessed and optimised in line with local community preferences for future coastline configuration.

One of our major findings is that there is no effect of degree of beach deterioration – characterised as loss of width and/or length of sandy beaches of between 5% and 100% - on respondents' willingness to pay for a management levy (although we note an exception for ***nay-leaners'*** preferences for Bay beaches). This finding suggests that respondents who agreed to pay a management levy were motivated to preserve sandy beaches in their current state irrespective of the severity of sand loss likely to occur as a result of coastal erosion. Respondents were willing to pay a levy to avoid even small losses (5% of the current sand volume). This is consistent with the economic theory that highlights general unwillingness to accept the loss (survey respondents typically assign a relatively higher value to an averted loss than to a potential gain of similar magnitude; Camerer (2005), Schmidt & Zank (2005)). Our results indicate that there is a strong 'preservationist' attitude, whereby survey respondents demonstrated a preference to maintain NSW beaches in their current quantity and condition. This can be considered a positive result in terms of sustainable financing perspective in that WTP spans the breadth of the coastline and is not merely reactionary to intense coastal risk or damage, but it presents difficulties in terms of strategic prioritisation.

However, other of our findings can assist with spatial prioritisation of coastal erosion management. These include our finding that respondents were willing to pay different levy amounts for different types of beaches (Iconic, Main, Bay and Surf). We also find differences in WTP amongst different populations segments (***yea-sayers, yea-leaners, nay-leaners, nay-sayers***). To the extent that these can be linked to socio-demographic characteristics (see Section 5.2), these can also be used to discriminate the value and beach preferences of a specific community or local government area in order to assist with spatial prioritisation.

We have used the time- and distance decay trends in WTP to infer the types of non-market values held by different (***nay-leaning***, ***yea-leaning*** and ***yea-saying***) respondent groups. We rely on the "pragmatic approach to measuring non-use values" published by Marre et al. (2015). Three survey segments (***nay-leaning***, ***yea-leaning*** and ***yea-saying***) reported some WTP for a levy with 50 years payment horizon. Given that for ~80% of those surveyed (those aged 35 or above) a 50 years payment horizon goes beyond their own life expectancy - currently 82.9 years for Australians (World Health Organization 2016)- we consider that respondents from each of these groups held some component of bequest value. We note there was no tendency towards increased WTP a levy for beaches close to a respondent's home for ***yea-saying*** and ***yea-leaning*** groups – suggesting this group holds existence value that is independent of the likelihood that they will visit a specific beach, either now or in the future. In contrast, **nay-sayers** demonstrated increased WTP for beaches close to home, pointing to some element of option value. An increased WTP for local beaches also has implications for scale at which levy might be implemented. It suggests that a levy that is administered locally will increase acceptability to a broader segment of the population (capturing ~35% of the population we identified as ***nay-leaning***).

The format of our survey makes it possible to estimate the aggregate willingness to pay for a proposed levy amount and duration for the preservation of a specific beach. An example is provided in Figure 17. For demonstration purposes, willingness to pay has been broken down by beach users and non-beach users (a common approach in valuation studies and one that is sometimes of interest to policymakers). WTP as presented in Figure 17 also takes into consideration whether an individual (user or non-user) is a ***yea-sayer***, ***yea-leaning***, ***nay-leaning*** or ***nay-sayer***. The example shows WTP arising from a levy amount of $5 for a period of 10 years to maintain the current condition of Austinmer beach in Wollongong LGA. This results in a total WTP value of $142,330 per annum, where $108,261 is derived from the beach users and $34,069 from non-users. Furthermore, it shows that out of the $108,261 for beach users that are yea-sayers, their WTP is $41,024, and the yea-leaning group are WTP $22,149. Nay-leaning has a total WTP of $45,088 and finally, for the nay-sayers, there is no WTP at any levy amount. The proportion of households belonging to each segment that is willing to pay the specific management levy is also given. We emphasise that the difference in WTP between users and non-users identified in Figure 17 emerges from differences in the relative proportion of survey segments (***yea-sayers, yea-learners***, ***nay-leaners***, ***nay-sayers***), rather than from differences in the underlying values *per se*. This highlights the value of our latent class segmentation approach and the benefits for interpretation it can provide over traditional models that treat these groups as separate and different entities.

**Figure 17**: A hypothetical example of calculating the total WTP value for a specific policy scenario

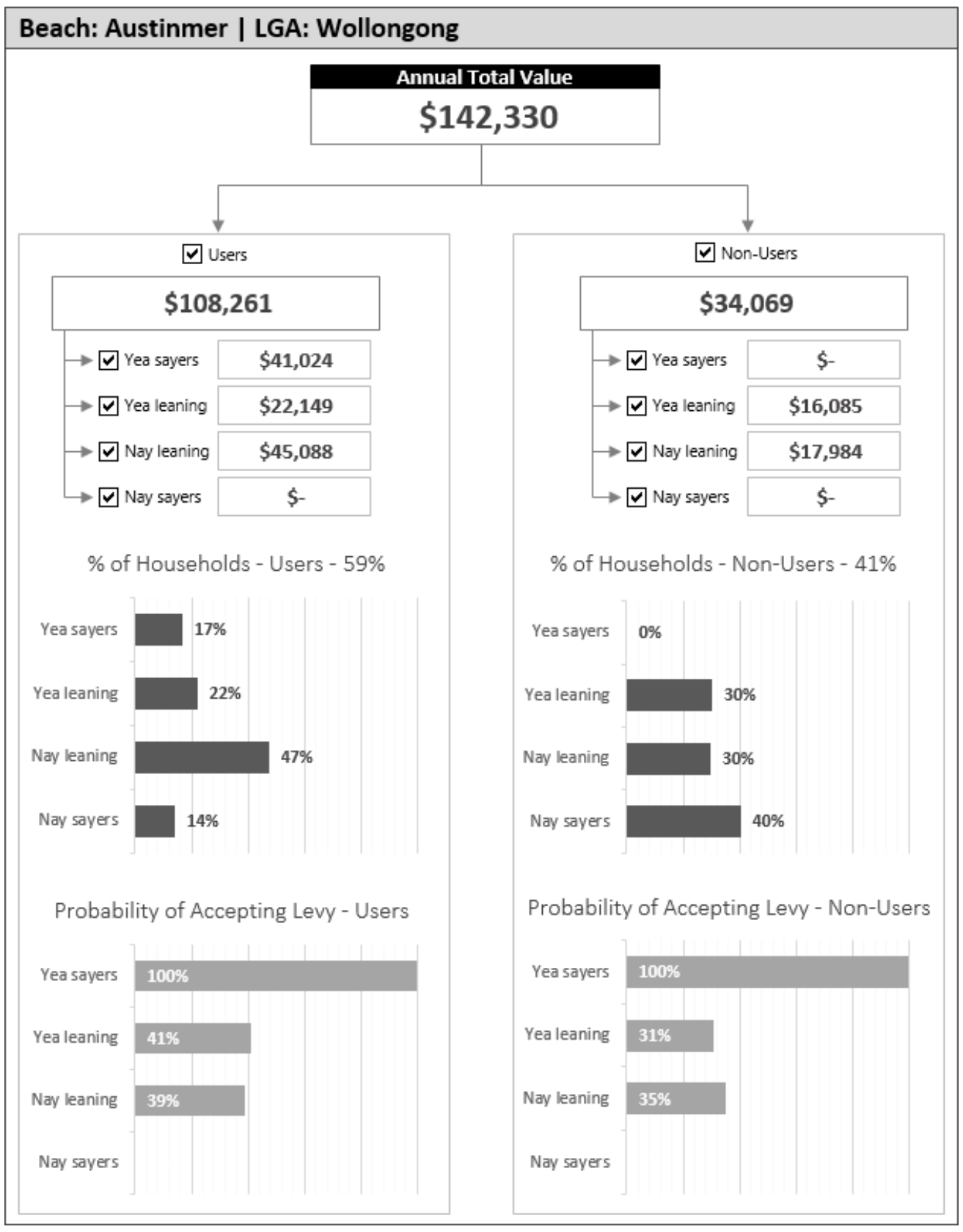

As a final note of the nature of non-market values for sandy beaches presented in this study, we highlight that it is not necessarily the case that ***nay-sayers*** hold a zero non-use value for sandy beaches. An alternative explanation is that they may have lodges a 'protest vote' about the proposed payment vehicle (annual levy) or about where responsibility for further investment in coastal protection lies – they may think that it should already be covered in their taxes. The size of this proportion of the population (35%) may be of concern if decision-makers are seeking to implement a mandatory levy to support coastal management. We recommend further research to identify respondents' motivations for nay-saying in order to determine if a more acceptable payment vehicle might be conceived.

# 7. CONCLUSIONS

Many authors have reported that the environmental preservation and management of natural or protected areas funds are insufficient and declining (Banhalmi-Zakar, Ware, Edwards, Kelly, & Becken, 2016; Baral, Stern, & Bhattarai, 2008; Dharmaratne, Sang, & Walling, 2000; Eagles, McCool, Haynes, Phillips, & Programme, 2002; Lindberg, 1998; Reynisdottir, Song, & Agrusa, 2008). Our study suggests there is WTP within the community to preserve sandy beaches in the face of future coastal erosion, raising the possibility of realising better funding arrangements for the preservation of coastal assets.

Our findings contribute to the current literature by providing significant empirical findings that coastal managers can benefit from in their decision making as well as investigating a new public funding mechanism. Moreover, from a methodological perspective, this study is innovative in using the standard latent class model in the treatment of strategic or protest voting in the form of "nay-saying" as well as "yea-saying" at the estimation stage rather than through elimination by the researcher prior to the estimation.